\pdfoutput=1 

\RequirePackage{fix-cm}
\documentclass[twocolumn]{svjour3}          
\smartqed  
\usepackage{graphicx}

\usepackage{subcaption}
\usepackage{varioref}

\usepackage{multirow}
\usepackage{boldline} 

\usepackage[hyphens]{url}
\usepackage[hidelinks]{hyperref}
\hypersetup{breaklinks=true}
\usepackage{xcolor}
\definecolor{light-gray}{gray}{0.95}

\usepackage{booktabs}
\usepackage{amsmath}
\usepackage{verbatim}

\begin{document}
\sloppy

\title{A calibration method structured on Bayesian Inference of the HCM speed-flow relationship for freeways and multilane highways and a temporal analysis of traffic behavior 
}

\titlerunning{A calibration method of the HCM speed-flow relationship and a temporal analysis of traffic behavior}        

\author{Gabriel Martins de Oliveira        \and
        Andre Luiz Cunha 
}


\institute{O. Gabriel \at
              Av. Trab. Sao Carlense, 400 - Parque Arnold Schimidt, Sao Carlos - SP, Brasil, 13566-590\\
              \email{gabrielmo83@usp.br}           
           \and
           A. Cunha \at
              Av. Trab. Sao Carlense, 400 - Parque Arnold Schimidt, Sao Carlos - SP, Brasil, 13566-590\\
              Tel: +55-16-3373-9597
              \email{alcunha@usp.br}           
}

\date{Received: date / Accepted: date}

\maketitle

\begin{abstract}
This paper presents a calibration method for the speed-flow model of the HCM 2016 for freeways and multilane highways allied to temporal analysis of traffic stream. The proposed method was developed using a sample of more than one million observations collected by 23 traffic sensors on four highways in the state of S{\~a}o Paulo. The method is structured on Bayesian inference and provided for each model parameters a probability distribution function. The free-flow speed and capacity presented a probability density function that approximates a Normal distribution. The segment in which the speed of traffic stream remain constant with the increase of the traffic flow is lower than described in HCM 2016, being in some cases close to zero. Along with the proposed calibration method an analysis of temporal variation is performed which shows a significant variation in traffic behavior for different periods. The free-flow speed, capacity and breakpoint distributions obtained through monthly and annual calibration were considered equal by means of Kolmogorov-Smirnov test, different for the model calibration coefficient.
         
\keywords{Traffic Engineering \and Calibration \and Traffic Behavior \and Bayesian Inference  \and Highway Capacity Model \and Freeways \and Multilane \and Temporal Variability}
\end{abstract}

\section{Introduction}
\label{secIntroducao}

Important decisions of investments on highways are significantly influenced by the results of analyses of one or more existing or planned segments that form the highway. The analyses are directly associated to traffic flow theories, which allow to describe the phenomenon related to traffic behavior. Thus, the traffic flow theories become indispensable in the implementation of models and tools for the planning and management of transport infrastructures \cite{pompigna2015}.

Traffic flow theories seek to describe the interaction between vehicles, drivers and road infrastructure \cite{klein2006}. Among the main characteristics is the presence of elements with behavioral, spatial and temporal variability, due to, respectively, the users of the system, the place under analysis and the study period. Temporal variability is related to the amount of data, data aggregation and cover period. One day presents high variance of traffic patterns that is represented mostly through the use of peak hour factor (PHF) \cite{roess2010}. On the same idea, different days of the week present different patters from others, as example on Thursday and Sunday. This temporal variability is also present in months and periods of years. 

Therefore, it is necessary to understand the implications that temporal variability has on traffic stream behavior. This study aims to calibrate the HCM speed-flow model for freeways and multilane highways trough Bayesian Inference and analyze the impact of perform it monthly and annually as well as to comprehend the inner relationship between the model parameters obtained through different periods. The option to use the HCM framework consist that it has been widely adopted worldwide as a main reference to estimate level of service, included in Brazil.

This paper is organized as follow. Firstly a literature review on the subjects is presented: mathematical modeling of the HCM speed-flow relationship and concepts used to perform calibration - Bayes inference. Then, traffic data used is presented. Next, the procedure to perform the calibration of the speed-flow relationship is describe as well as its results, followed by a comparison with other methods. Then an analysis of temporal variability is performed. Finally conclusions and recommendations are presented.

\section{Literature Review}

\subsection{Mathematical modeling of the HCM speed-flow relationship}

The HCM 2010 and 2016 editions use the same mathematical modeling as the HCM 2000 to describe the speed-flow relationship of traffic stream for multilane and freeways \cite{hcm2000,hcm2010,hcm2016}. The model consists of five parameters which represent a curve with two segments that vary according to traffic flow, expressed by:

\begin{equation}
\left \{
{ \begin{array}{lcl}
	q \leq {b}_p  & , & \textit{u} = {\textit{u}}_f \\
	{b}_p < q \leq {q}_c & , & \textit{u} = {\textit{u}}_f - \bigg ({\textit{u}}_f{} - \frac {{q}_c}{{k}_c} \bigg) \cdot \bigg(\frac{q - {b}_p}{{q}_c - {b}_p} \bigg)^\alpha
	\end{array}} \right. 
\label{eq:8}
\end{equation}

Where:

\begin{itemize}
	\item $q$: Traffic flow rate; 
	\item $\textit{u}$: Average speed;	 
	\item ${\textit{u}}_f$: Free-flow speed;
	\item ${q}_c$: Flow at capacity; 
	\item ${k}_c$: Density at capacity;	
	\item \textit{bp}: Breakpoint; and 
	\item $\alpha$: Calibration coefficient.		
\end{itemize}

\begin{figure}[!htbp]
	\centering
	\begin{minipage}[t]{0.4\textwidth}
		\centering		
		\includegraphics[width=\textwidth]{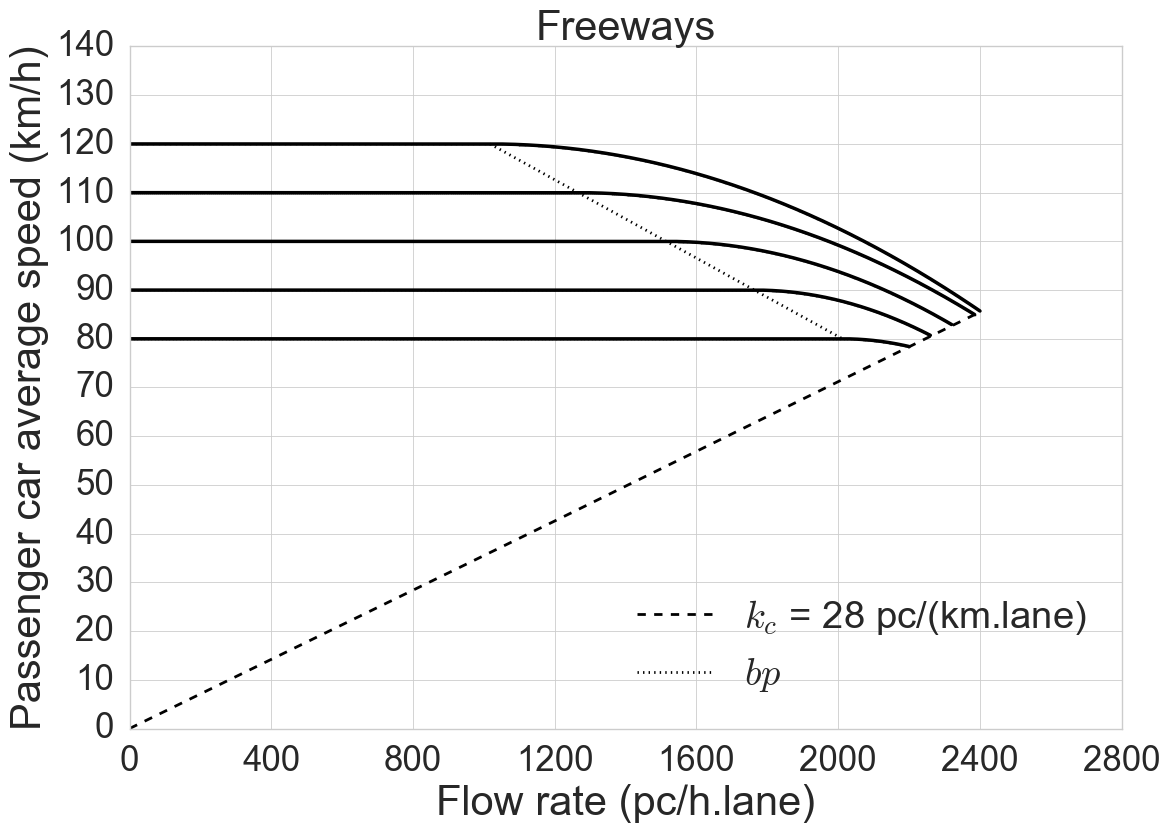}
		\subcaption{}
		\label{fig:sub:freeway}
	\end{minipage}
	\hspace{\fill}
	\begin{minipage}[t]{0.4\textwidth}
		\centering		
		\includegraphics[width=\textwidth]{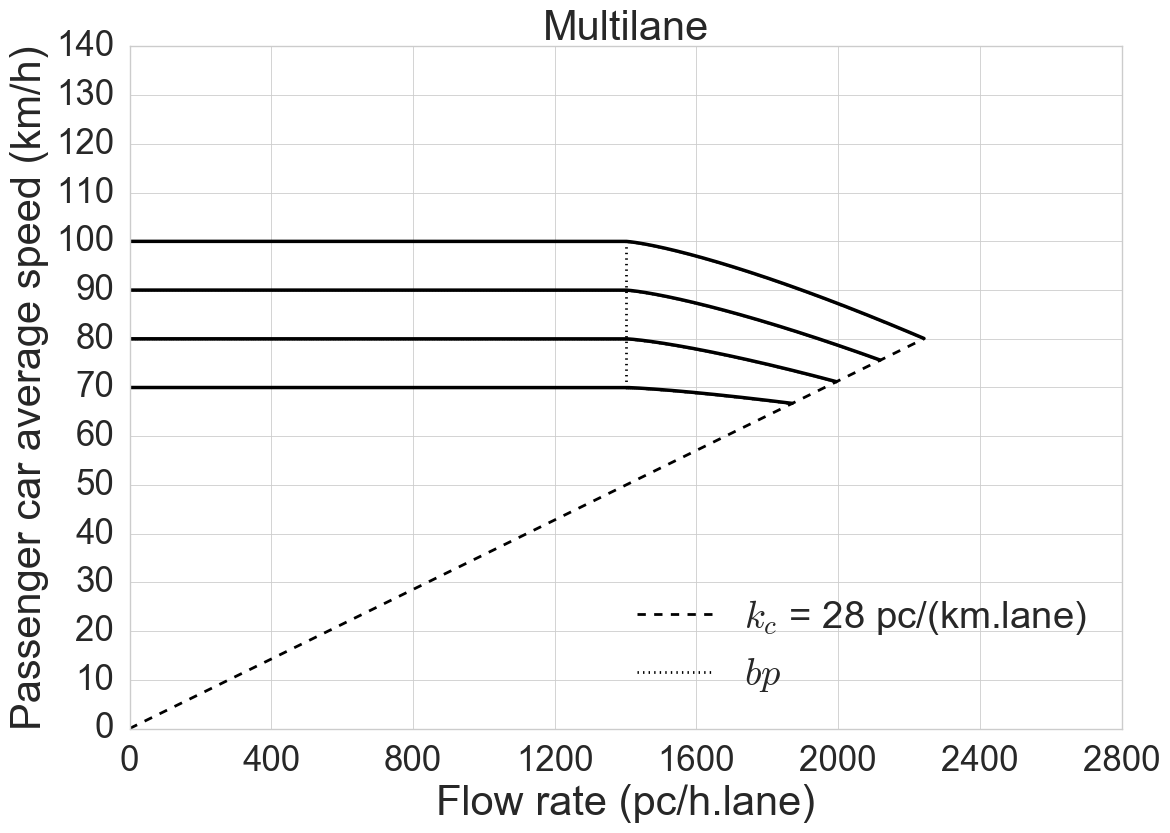}
		\subcaption{}
		\label{fig:sub:multilane}
	\end{minipage}
	\caption{HCM 2016 Speed-flow curves for (\subref{fig:sub:freeway}) Freeway and (\subref{fig:sub:multilane}) Multinale}
	\label{fig:hcm_curves}
\end{figure}

The first segment is based on the assumption that there is a speed threshold, characterized by the free-flow speed (${\textit{u}}_f$), where the speed remains constant with increasing traffic flow to a transition point (\textit{bp}). The second one consists of a convex curve anchored in \textit{bp} and density at capacity ($ kc$) \cite{roess2011}. This segment is characterized by the decrease of the average speed with the increase of traffic flow. Figure~\ref{fig:hcm_curves} shows the families of curves for freeways and multilane of the HCM 2016. Both freeways and multilane used the same mathematical modeling to represent traffic behavior being the main differences the values used for the breakpoint \cite{hcm2010,hcm2016}. 

\subsection{Bayesian inference}

In science, all available information about a natural phenomenon is always incomplete, and knowledge about nature is necessarily probabilistic. Thus, statistical inference is based on a theory of probability \cite{gregory2005}.

Over time, two main approaches of statistical inference were developed based on different definitions of probability: the classic, also called frequentist, and Bayesian. In classical statistics, the concept of probability of an event is defined as the relative frequency of occurrence of this event after successive repetitions, whereas, in Bayesian statistics, the probability of an event is the degree of knowledge about it \cite{vanderplas2014}.

\subsubsection{Bayes Theorem}

Bayesian Inference is a probabilistic statistical method based on the Bayes' theorem which consists of the probability of a hypothesis varying in time as more information about it is added \cite{downey2013}.

\begin{equation}
		p(H|D) = \frac{p(H)p(D|H)}{p(D)}
\label{eq:10}
\end{equation}

\begin{itemize}
	\item $p(H)$: is the probability of the hypothesis before data observation; called prior probability;
	\item $P(D)$: it is the probability of the data under any circumstances; called the normalization constant;	 
	\item $p(D|H)$: is the probability of the data under the hypothesis in question; called the likelihood function; and
	\item $p(H|D)$: is the probability to be calculated after new data are observed; called posterior probability.
\end{itemize}

In short, Bayesian analysis is a statistical method that makes inferences about unknown quantities as model parameters, joining previous knowledge about the subject with existing evidence represented by a set of data. Thus, a Bayesian model can be interpreted in two parts: a likelihood function, which describes the distribution of the data on the subject in question and a prior distribution, which reflects the prior knowledge about the subject, regardless of data observation. From the likelihood function, the prior distribution and the data observation, the prior knowledge is updated by Bayes' theorem, resulting in the estimation of the posterior distribution \cite{chen2009}.

In theory, posterior distributions contain all the information necessary to make inferences about the subject they represent. However, in practice, posterior distributions are difficult to estimate, especially analytically and for complex problems, since they involve the calculation of multiple integrals. Thus, to obtain estimates of posterior distributions, most Bayesian analyzes are structured in sophisticated computational methods, such as simulation methods based on Markov chain Monte Carlo \cite{gamerman2006,brooks2011}.

\subsubsection{Markov chain Monte Carlo}

The MCMC methods correspond to simulation methods that, when correctly defined and implemented, make it possible to perform successive samplings of posterior distributions. The Monte Carlo step of the MCMC methods consist in estimating the properties of distributions by analyzing random samples of these distributions while the Markov chains are based on the concept that random samples are generated by specific sequential processes in which the generation of a new sample depends on the previous sample.

The convergence of MCMC process is difficult to determine precisely and several methods have been proposed. The Gelman and Rubin test is one of the most popular in the literature \cite{cowles1996}. The test evaluates the convergence of the MCMC by the analysis of multiple Markov chains. To do so, it is based on the principle that if multiple chains converge, then they must be similar to each other; and if they are not, the chains do not converge.

To verify the similarity, the test performs a variance analysis, in which it compares the variance of the estimated parameters of the model between the same Markov chain and between the multiple chains. A large difference between the variances indicates the non-convergence of the MCMC method \cite{brooks1998}.

The test statistic is based on an internal parameter, $\hat{R}$, which represents the level of association between the variances of a chain and multiple chains. The authors of the test suggest that it is possible to safely assume that the MCMC has achieved convergence when all parameters of the model have $\hat{R} <1.1$. 

\section{Data collection}

In order to support the calibration of the HCM speed-flow relationship for freeways and multilane highways, a sample with more than one million observations was used, collected by 23 traffic sensors on four highways at S{\~a}o Paulo state (SP-280, SP-348, SP-270, and SP-021). The data were collected through inductive loops installed in each traffic lane between 07/01/2011 and 06/01/2012 and consisted of number of vehicles and average speed (for passenger cars and heavy vehicles). For 8 of the 23 sites, data were available in 6-minutes intervals while the 15 other sites, data were in 5-minute intervals. According to the HCM, all sites in the sample are basic segments and could be classified either as freeways (expressways with controlled access) or as divided multilane highways (expressways without controlled access).

Sites in the sample were also classified as ``rural'' or ``urban'', according to abutting land use. ``Rural expressways'' consist on highways isolated from the local road network, producing mostly longer trips; ``urban expressways'' are those with greater integration with the local area and carrying a significant number of local trips. Road characteristics were obtained through \cite{andrade2011} and consist of highway type, abutting land use, number of lanes, post speed per vehicle type, vertical alignment, horizontal alignment and grade. Table~\ref{tab:caracteristica_fisica_sensores} presents a summary of each site used in the study.

\begin{table*}[htbp]
	\centering
	\caption{Characteristics of road segments}
	\resizebox{\linewidth}{!}{%
		\begin{tabular}{lllllcccccc}
			\toprule
			\multicolumn{1}{p{4.215em}}{Highway} &
			\multicolumn{1}{p{4.215em}}{km} & \multicolumn{1}{p{6.145em}}{Direction} & \multicolumn{1}{p{6.0em}}{Highway type} & \multicolumn{1}{p{7.0em}}{Abutting land use} & \multicolumn{1}{p{3.57em}}{Number of lanes} & \multicolumn{1}{p{6.4em}}{Post speed automobile (km/h)} &
			\multicolumn{1}{p{6.0em}}{Post speed heavy vehicles (km/h)} &
			\multicolumn{1}{p{5em}}{Vertical alignment (m/km)} & \multicolumn{1}{p{4.43em}}{Horizontal alignment (°/km)} & \multicolumn{1}{p{3.2em}}{Grade (\%)} \\
			\midrule
			SP-348 & 32,0  & North & Freeway & Rural & 4     & 120   & 90    & 37,2  & 42,3  & 3,5 \\
			SP-348 & 32,0  & South   & Freeway & Rural & 4     & 120   & 90    & 37,2  & 42,3  & -3,5 \\
			SP-348 & 50,0  & North & Freeway & Rural & 3     & 120   & 90    & 41,4  & 12,6  & -3,0 \\
			SP-348 & 50,0  & South  & Freeway & Rural & 3     & 120   & 90    & 41,4  & 12,6  & 3,0 \\
			SP-348 & 65,0  & North & Freeway & Rural & 3     & 120   & 90    & 20,9  & 21,8  & -2,0 \\
			SP-348 & 65,0  & South  & Freeway & Rural & 3     & 120   & 90    & 20,9  & 21,8  & 2,0 \\
			SP-348 & 87,0  & North & Freeway & Rural & 3     & 120   & 90    & 25,0  & 21,6  & 2,5 \\
			SP-348 & 87,0  & South  & Freeway & Rural & 3     & 120   & 90    & 25,0  & 21,6  & -2,5 \\
			SP-280 & 16,0  & West & Multilane & Urban & 3     & 100   & 90    & 20,0  & 35,4  & 2,0 \\
			SP-280 & 22,4  & East & Freeway & Urban & 3     & 100   & 90    & 22,0  & 61,2  & -1,0 \\
			SP-280 & 22,4  & West & Freeway & Urban & 3     & 100   & 90    & 22,0  & 61,2  & 1,0 \\
			SP-280 & 29,5  & East & Freeway & Urban & 3     & 120   & 90    & 43,6  & 52,1  & 2,0 \\
			SP-280 & 37,0  & East & Multilane & Rural & 3     & 120   & 90    & 47,4  & 16,3  & 5,0 \\
			SP-280 & 37,0  & West & Multilane & Rural & 3     & 120   & 90    & 47,4  & 16,3  & -5,0 \\
			SP-280 & 51,9  & East & Multilane & Rural & 3     & 120   & 90    & 33,5  & 20,1  & 4,5 \\
			SP-280 & 51,9  & West & Multilane & Rural & 3     & 120   & 90    & 33,5  & 20,1  & -4,5 \\
			SP-280 & 59,6  & East & Multilane & Rural & 3     & 120   & 90    & 42,7  & 13,6  & -2,0 \\
			SP-280 & 59,6  & West & Multilane & Rural & 3     & 120   & 90    & 42,7  & 13,6  & -5,0 \\
			SP-280 & 75,9  & East & Multilane & Rural & 3     & 120   & 90    & 33,0  & 16,3  & 4,0 \\
			SP-280 & 75,9  & West & Multilane & Rural & 3     & 120   & 90    & 33,0  & 16,3  & -4,0 \\
			SP-270 & 39,9  & East & Multilane & Urban & 2     & 80    & 80    & 53,2  & 78,2  & 6,5 \\
			SP-021 & 18,3  & North & Multilane & Urban & 4     & 100   & 80    & 5,4   & 18,5  & -0,5 \\
			SP-021 & 22,3  & South & Multilane & Urban & 4     & 100   & 80    & 14,4  & 26,2  & 2,5 \\
			\bottomrule
		\end{tabular}%
	}
	\label{tab:caracteristica_fisica_sensores}%
\end{table*}%

Furthermore, data treatment and cleaning process were performed with the purpose of removing observations considered inappropriate from a model calibration perspective through the guidelines made by the authors that developed the manual. This procedure was performed by filtering the original data through three main criteria: traffic flow only composed by passenger car, traffic under normal operations condition, and only containing observations in which the free flow regime applies. To accomplish the first two criteria, only data collected by the sensors on the leftmost lane of each site were used. Additionally, observations with presence of any heavy vehicles were discarded (${P}_t >$ 0~\%) in order to not consider the heavy-vehicle factor impact. Also, were discarded observations considered with anomalies, such as negative average speed, traffic flow equal to zero and associated average speed different than zero and average speed bigger than 180~km/h. 

\begin{figure}[!htbp]
	\centering
	\begin{minipage}[t]{0.4\textwidth}		
		\centering
		\includegraphics[width=\textwidth]{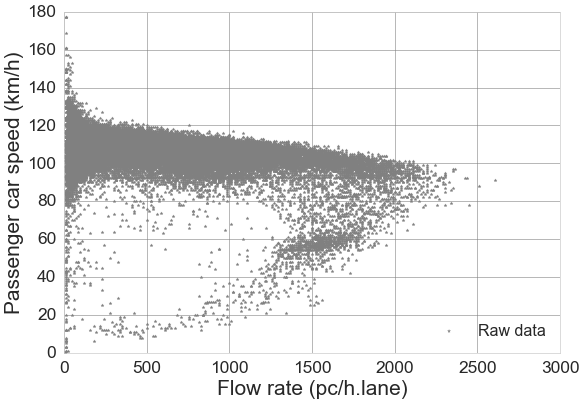}
		\subcaption{}
		\label{fig:sub:dados_brutos_calibracao}
	\end{minipage}
	\hspace{\fill}
	\begin{minipage}[t]{0.4\textwidth}
		\centering
		\includegraphics[width=\textwidth]{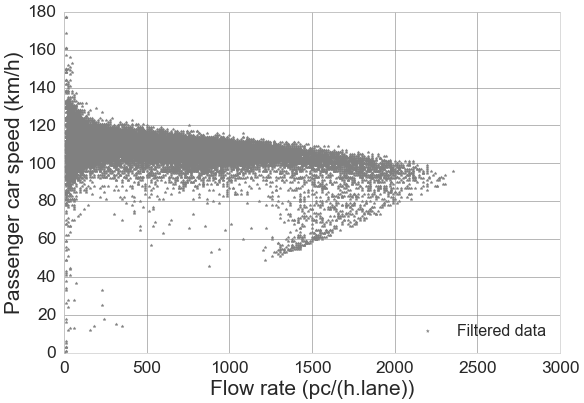}
		\subcaption{}
		\label{fig:sub:dados_filtrados_calibracao}
	\end{minipage}
	\caption{Observations of traffic flow and speed for SP-280, km 51,9 east (\subref{fig:sub:dados_brutos_calibracao}) Before treatment and (\subref{fig:sub:dados_filtrados_calibracao}) After treatment}
	\label{fig:dados}
\end{figure}

The third filtering criteria was applied by splitting free flow and congested regime through the threshold of density at capacity. To this extent, it was assumed a density at capacity for both ``rural'' and ``urban'' highways, respectively, of 26 and 25~pc/(km.lane), as indicated in \cite{andrade2014} for highways located at S{\~a}o Paulo state in Brazil. Figure~\ref{fig:dados} shows data before and after treatment for sensor located on SP-280 km 51,9 east. After the treatment the sensor with more observations for the period of one year had 81.232 and with less had 27.358 observations.

\section{Calibration of speed-flow relationship}

The method used for the calibration of the speed-flow relationship of the HCM model for freeways and multilane highways through Bayesian inference consist of the following steps: data collection and treatment of the database; segments classification and definition of density at capacity; definition of prior distributions of model parameters; likelihood function and Markov chain Monte Carlo; estimation of posterior distribution of the model parameters and definition of a representative value for the posterior distribution.

\subsection{Segments classification}

It was chosen to classify the segments according to the highway abutting land use, in ``rural'' and ``urban'' areas, as proposed by \cite{andrade2014}. They describe that it makes little sense to divide the highways of the state of S{\~a}o Paulo in freeways and multilane, as recommended by HCM. Instead, they proposed a division between ``rural'' and ``urban'' highways as the main differences between freeways and multilane in the state of S{\~a}o Paulo refer to access control and the density of uncontrolled access to the highway. This characteristics differ from freeways and multilane in the US, which have distinct physical characteristics, such as intersections and occasional traffic signals.

The characteristics related to the highways of the state of S{\~a}o Paulo, access control and density of uncontrolled access, consist of typical attributes associated to the highway abutting land use. The density at capacity for rural and urban segments was determined as 26 and 25~pc/(km.lane), respectively. 

\subsection{Prior distribution}  

The distributions of the parameters, ${\textit{u}}_f$, ${q}_c$, \textit{bp}, $\alpha$, of the speed-flow relationship of the HCM 2016 model were determined as uniform. The lower and upper range of each parameter is shown in Table~\ref{tab:distribuicao_priori}.

\begin{table}[htbp]
	\centering
	\caption{Prior distribution of HCM model parameters}
	\resizebox{\columnwidth}{!}{%
	\begin{tabular}{p{6.215em}lll}
		\toprule
		Parameter & \multicolumn{1}{p{9.785em}}{Type of distribution} & \multicolumn{1}{p{7.145em}}{lower band} & upper band \\
		\midrule
		${\textit{u}}_f$   & Uniform & 0 km/h & \multicolumn{1}{p{7.57em}}{160 km/h} \\
		${q}_c$    & Uniform & 0 pc/h.ln & \multicolumn{1}{p{7.57em}}{2.800 pc/h.ln} \\
		\textit{bp}    & Uniform & 0 pc/h.ln & \multicolumn{1}{p{7.57em}}{2.000 pc/h.ln} \\
		$\alpha$  & Uniform & 1     & 3 \\
		\bottomrule
	\end{tabular}%
	}
	\label{tab:distribuicao_priori}%
\end{table}%

The option to use a uniform distribution to represent the model parameters is to choose a more objective prior distribution possible, where it is assumed that the behavior of the parameters is not known \cite{berger2006,camdavison2015}. The uniform distribution has the characteristic that all values located into the lower and upper bound interval present the same probability of occurrence.

\subsection{Likelihood function} 

\begin{figure}[h]
	\centering
	\includegraphics[width=0.4\textwidth]{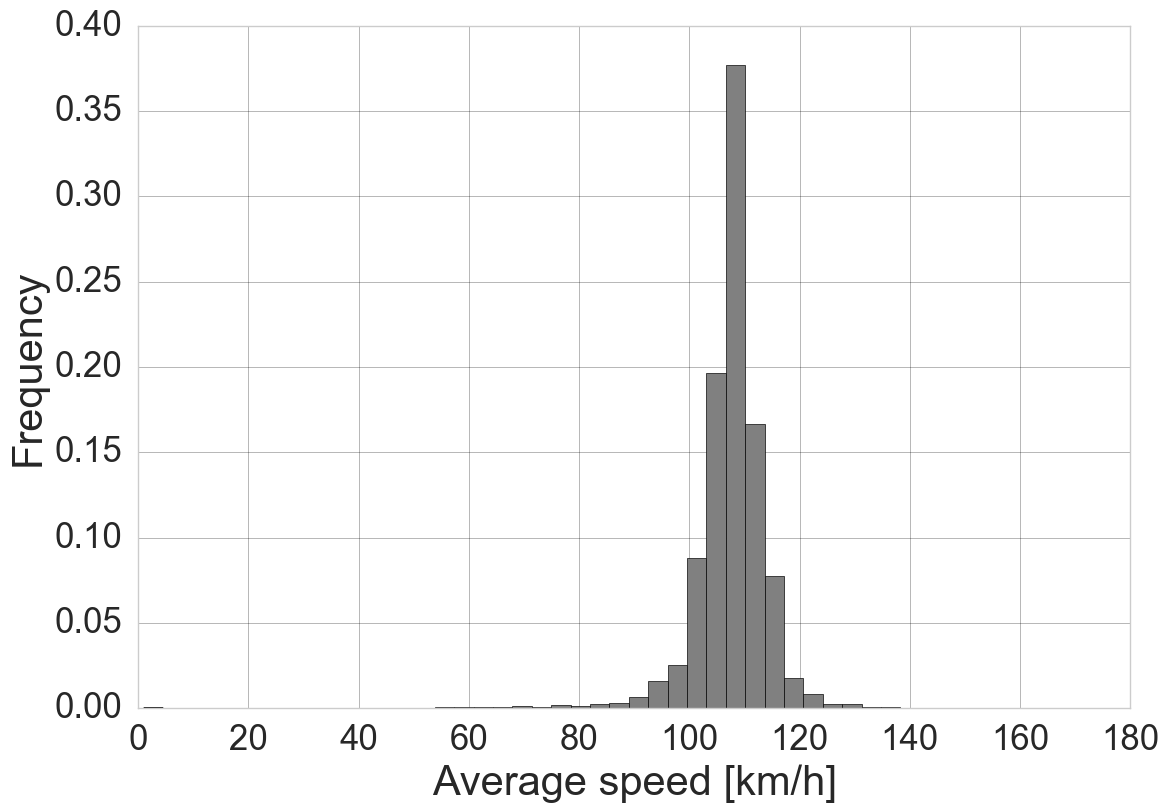}
	\caption{Probability density function of average speed for SP-280, km 51,9 east}
	\label{fig:funcao_verossimilhanca}
\end{figure}

The likelihood function was determined as a Normal distribution. For this, the probability density function (PDF) of the variable average speed of traffic flow was analyzed. The distribution parameters, mean and standard deviation, were obtained from the processed data. Figure~\ref{fig:funcao_verossimilhanca} shows the probability density function of the average speed of traffic flow for the sensor located on SP-280 km 51,9 east with filtered data.

\subsection{Markov chain Monte Carlos configuration} 

The MCMC parameters were chosen by the analysis of the posterior distribution, which involved the convergence of MCMC process and data processing time. For that, it was used 3 chains in MCMC with the same number of iterations and different start values. The reason to use different MCMC configurations was to perform the Gelman and Rubin test, which aims to investigate the convergence of MCMC process. 

\subsection{Results for the calibration} 

For the period of one year, the estimation of the parameters of the speed-flow relationship of the HCM model was performed for the 23 collection points, 15 of which reached the highway capacity. Altogether 16 segments were classified as rural and 7 as urban highways. In the results were evaluated: posterior distributions of parameters; MCMC convergence process; and comparison between values obtained from the proposed method with HCM 2016 \cite{hcm2016} and Andrade and Setti \cite{andrade2014}.

\subsubsection{Posterior distribution}

\begin{figure}[!htbp]
	\centering		
	\begin{minipage}[t]{0.4\textwidth}
		\centering		
		\includegraphics[width=\textwidth]{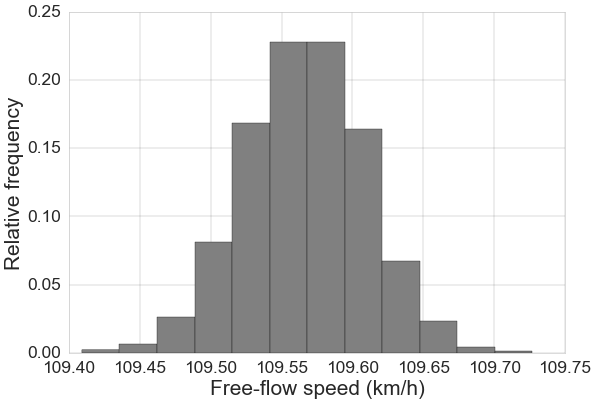}
		\subcaption{}
		\label{fig:sub:subd1}
	\end{minipage}
	\hspace{\fill}
	\begin{minipage}[t]{0.4\textwidth}
		\centering		
		\includegraphics[width=\textwidth]{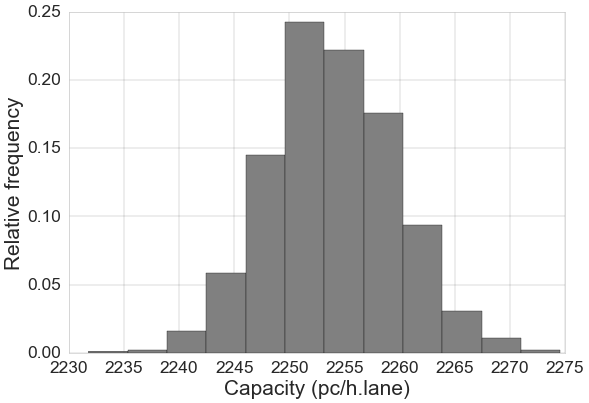}
		\subcaption{}
		\label{fig:sub:subd2}
	\end{minipage}
	\hspace{\fill}
	\begin{minipage}[t]{0.4\textwidth}
		\centering		
		\includegraphics[width=\textwidth]{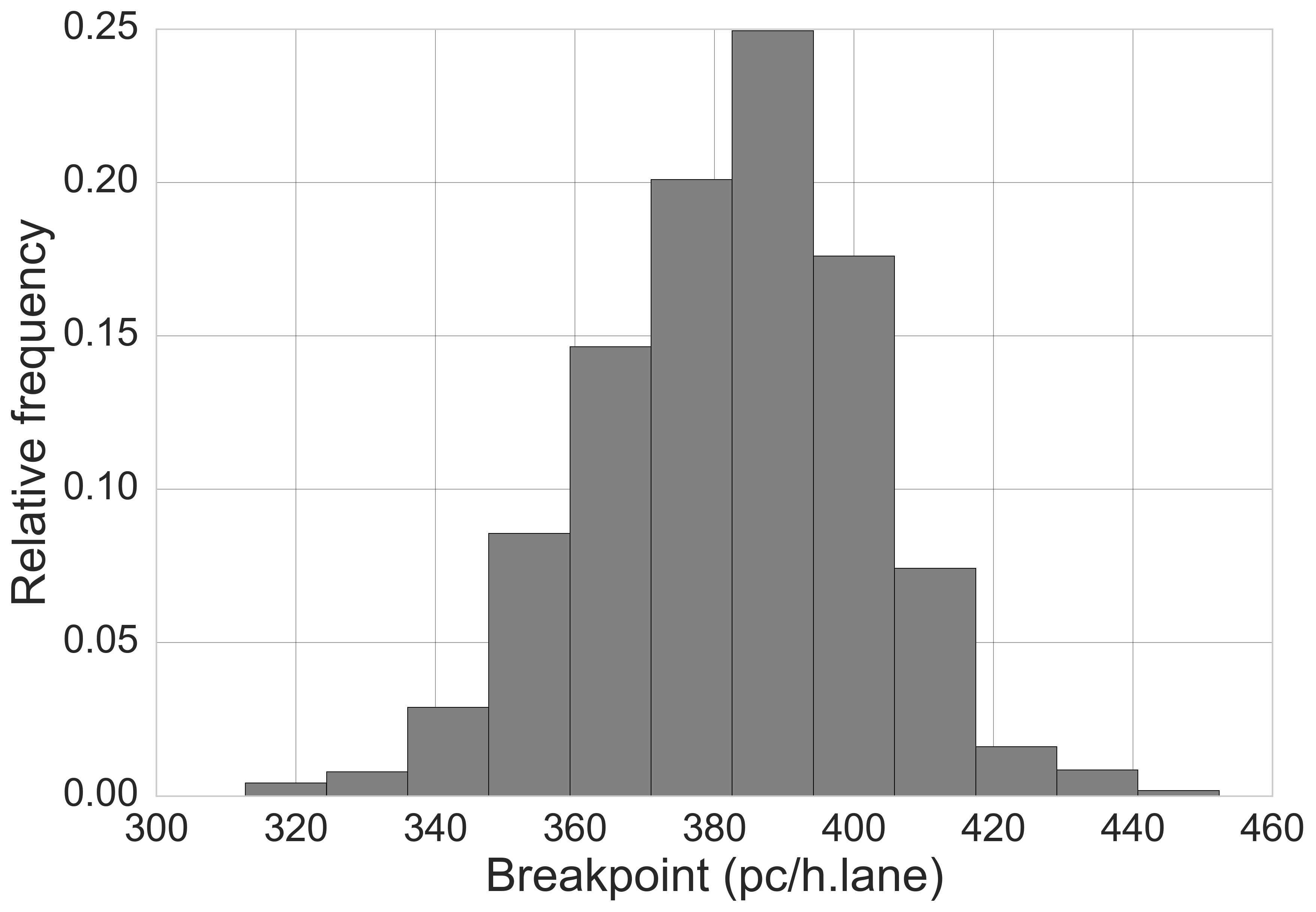}
		\subcaption{}
		\label{fig:sub:subd3}
	\end{minipage}
	\hspace{\fill}
	\begin{minipage}[t]{0.4\textwidth}
		\centering		
		\includegraphics[width=\textwidth]{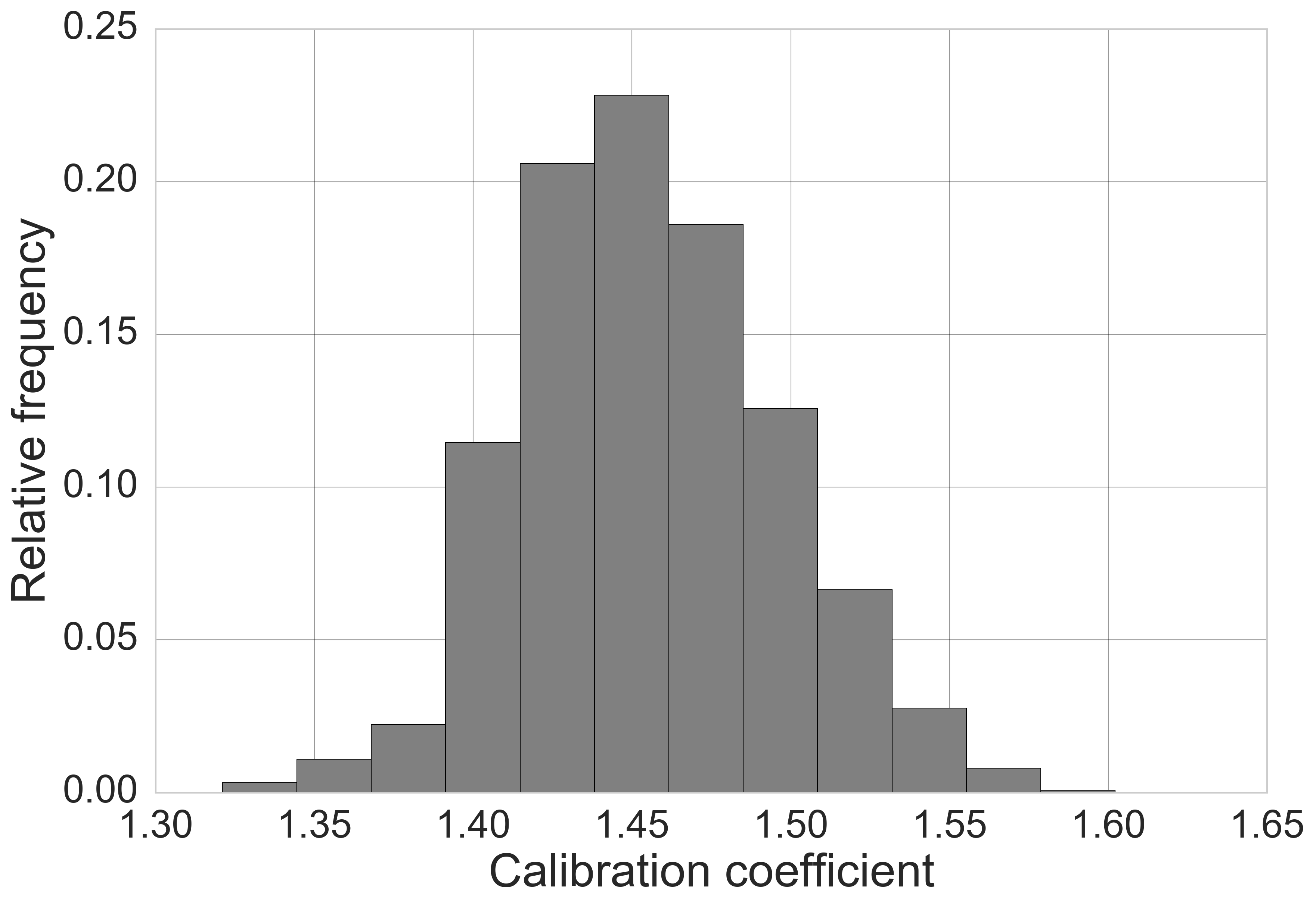}
		\subcaption{}
		\label{fig:sub:subd4}
	\end{minipage}
	\caption{Histograms for the proposed calibration method, HCM 2016 \cite{hcm2016} and Andrade and Setti \cite{andrade2014} for (\subref{fig:sub:subd1}) Free-flow speed (\subref{fig:sub:subd2}) Capacity (\subref{fig:sub:subd3}) Breakpoint and (\subref{fig:sub:subd4}) Calibration coefficient}
	\label{fig:distribuicao_posterior_calibracao}
\end{figure}

Figure~\ref{fig:distribuicao_posterior_calibracao} shows the posterior distributions of ${\textit{u}}_f$, ${q}_c$, \textit{bp} and $\alpha$, referring to SP-280 km 51,9, east. The mean of the distribution was used as the representative value of the posterior distributions, which in this case, corresponds to 109.6~km/h for ${\textit{u}}_f$, 2254~pc/h.lane for ${q}_c$, 383~pc/h.lane for \textit{bp} and 1.45 for $\alpha$. As a comparison, the values obtained with HCM 2016 \cite{hcm2016} for the same location were 109.8~km/h for free-flow speed, 2300~pc/h.lane for capacity, 1400~pc/h.lane for breakpoint, and 1.31 for calibration coefficient.

Noteworthy is the difference in the breakpoint values, which correspond to more than one thousand pc/h.lane regarding to the mean of the distribution and close to one thousand in relation to the distribution maximum value. It is also highlighted that for SP-280 km 51,9, east the HCM 2016 \cite{hcm2016} free-flow speed, capacity and calibration coefficient have values that are close to their respective distributions obtained through the proposed method.  

The shape of parameters distribution for all collection points used in the study range from Normal to Exponential distributions. Free-flow speed and capacity presented only Normal distribution, while \textit{bp} and $\alpha$ presented mostly Normal distribution, but in some cases, a Exponential distribution was detected. For \textit{bp} and $\alpha$, this occurred, respectively, 4 and 6 times in a total of the 23 collection points used in the study.  

\subsubsection{MCMC convergence}

The MCMC analysis was based on the verification of the MCMC convergence and the data processing time. For the convergence analysis, the autocorrelation and the variation of samples values were verified by means of a graphical analysis allied to the statistical test of Gelman and Rubin. Table~\ref{tab:config_cadeia_mcmc_calibracao} presents the settings used in the process, where Burn in consists of discarding the initial iterations while Thin corresponds to the discard of iterations at successive intervals.

\begin{table}[htbp]
	\centering
	\caption{Markov chain Monte Carlo configurations}
	\begin{tabular}{lccc}
		\toprule
		\multicolumn{4}{c}{MCMC} \\
		\midrule
		\multicolumn{1}{c}{Configuration} & Iterations & Burn in & Thin \\
		\midrule
		1     & 20.000 & 0     & 0 \\
		2     & 30.000 & 10.000 & 0 \\
		3     & 50.000 & 30.000 & 0 \\
		\bottomrule
	\end{tabular}%
	\label{tab:config_cadeia_mcmc_calibracao}%
\end{table}%

\begin{figure}[!htbp]
	\begin{minipage}[t]{0.2455\textwidth}		
		\centering
		\includegraphics[width=\textwidth]{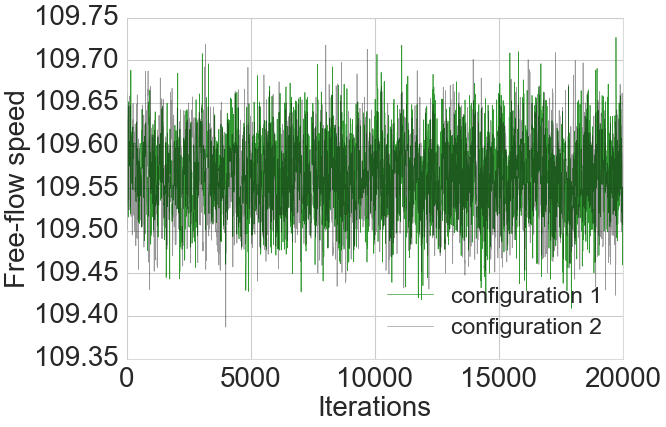}
		\subcaption{}
		\label{fig:sub:suba1}
	\end{minipage}
	\hspace{\fill}
	\begin{minipage}[t]{0.2255\textwidth}
		\centering
		\includegraphics[width=\textwidth]{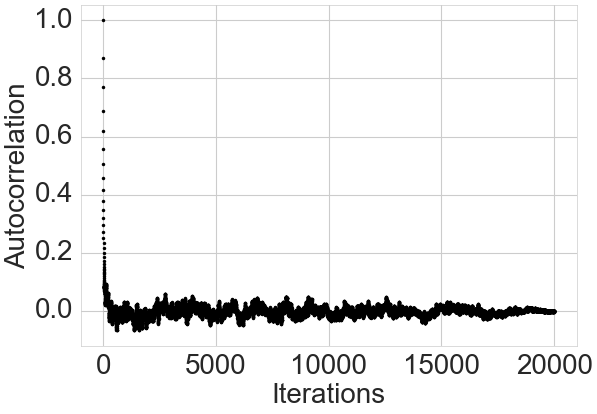}
		\subcaption{}
		\label{fig:sub:suba2}
	\end{minipage}
	\caption{Samples values (\subref{fig:sub:suba1}) and autocorrelation (\subref{fig:sub:suba2}) in function of number of iterations for ${\textit{u}}_f$}
	\label{fig:convergencia_velocidade_fluxo_livre}
\end{figure}

\begin{figure}[!htbp]
	\begin{minipage}[t]{0.2455\textwidth}		
		\centering
		\includegraphics[width=\textwidth]{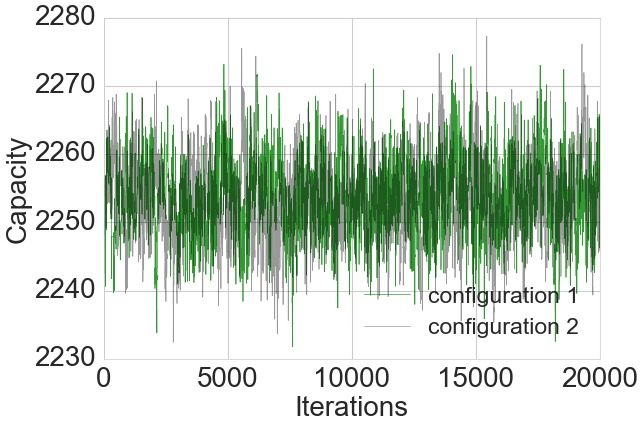}
		\subcaption{}
		\label{fig:sub:suba3}
	\end{minipage}
	\hspace{\fill}
	\begin{minipage}[t]{0.2255\textwidth}
		\centering
		\includegraphics[width=\textwidth]{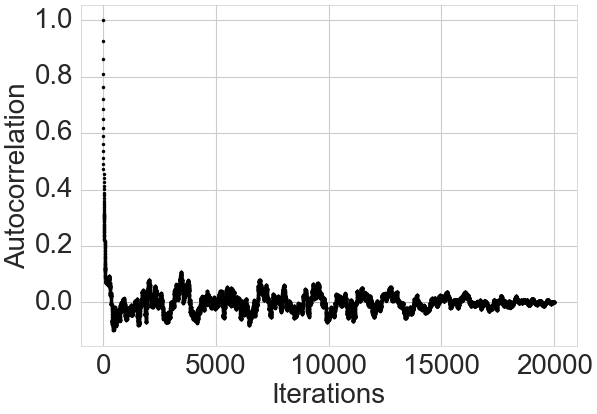}
		\subcaption{}
		\label{fig:sub:suba4}
	\end{minipage}
	\caption{Samples values (\subref{fig:sub:suba3}) and autocorrelation (\subref{fig:sub:suba4}) in function of number of iterations for ${q}_c$}
	\label{fig:convergencia_capacidade}
\end{figure}

\begin{figure}[!htbp]
	\begin{minipage}[t]{0.2455\textwidth}		
		\centering
		\includegraphics[width=\textwidth]{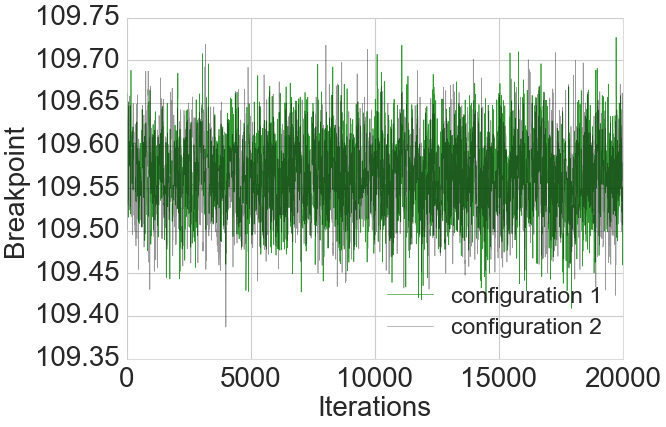}
		\subcaption{}
		\label{fig:sub:suba5}
	\end{minipage}
	\hspace{\fill}
	\begin{minipage}[t]{0.2255\textwidth}
		\centering
		\includegraphics[width=\textwidth]{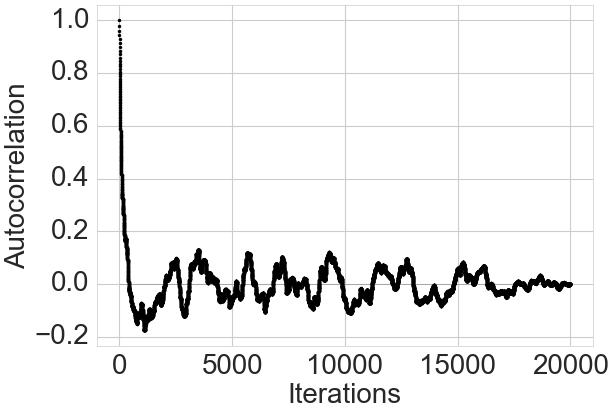}
		\subcaption{}
		\label{fig:sub:suba6}
	\end{minipage}
	\caption{Samples values (\subref{fig:sub:suba5}) and autocorrelation (\subref{fig:sub:suba6}) in function of number of iterations for \textit{bp}}
	\label{fig:convergencia_ponto_transicao}
\end{figure}

\begin{figure}[!htbp]
	\begin{minipage}[t]{0.2455\textwidth}		
		\centering
		\includegraphics[width=\textwidth]{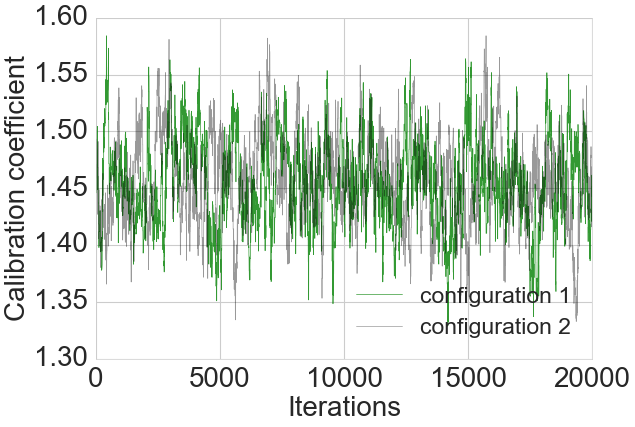}
		\subcaption{}
		\label{fig:sub:suba7}
	\end{minipage}
	\hspace{\fill}
	\begin{minipage}[t]{0.2255\textwidth}
		\centering
		\includegraphics[width=\textwidth]{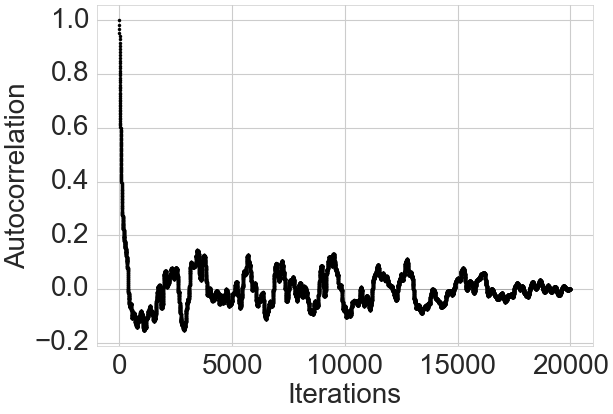}
		\subcaption{}
		\label{fig:sub:suba8}
	\end{minipage}
	\caption{Samples values (\subref{fig:sub:suba7}) and autocorrelation (\subref{fig:sub:suba8}) in function of number of iterations for $\alpha$}
	\label{fig:convergencia_alfa}
\end{figure}

Figures~\ref{fig:convergencia_velocidade_fluxo_livre}, ~\ref{fig:convergencia_capacidade}, ~\ref{fig:convergencia_ponto_transicao} and ~\ref{fig:convergencia_alfa} show the variation of sample values generated by the MCMC configurations and the autocorrelation, respectively, for ${\textit{u}}_f$, ${q}_c$, \textit{bp} and $\alpha$  referring to SP-280 km 51,9, east.

Figures~\ref{fig:convergencia_velocidade_fluxo_livre}, ~\ref{fig:convergencia_capacidade}, ~\ref{fig:convergencia_ponto_transicao} and ~\ref{fig:convergencia_alfa} show for all the MCMC configurations used that according to the increase in the number of iterations the autocorrelation decreases, indicating a convergence of the process. With the decrease in autocorrelation and the increase in the number of iterations, the variation of the parameter values vary around a well delimited area, pointing to a convergence of MCMC. It is highlighted that the larger the delimited area, the greater the credible region of the parameter, and the smaller the area, the lower the credible region, which graphically is represented by the area between the upper and lower limits of the speed-flow curves estimated by the proposed method.

In addition to the graphical analysis, the Gelman and Rubin test was applied, which provides indications of the convergence of the MCMC process by means of a statistical method based on the analysis of variance between different MCMC chains. For all collection point used in the study, all the parameters estimated presented Gelman and Rubin values $\hat{R}$ less than 1.1, which corroborates the convergence of MCMC process.

\subsection{Comparison with other methods}

\begin{figure}[!htbp]
	\begin{minipage}[t]{0.5\textwidth}		
		\centering
		\includegraphics[width=0.995\textwidth]{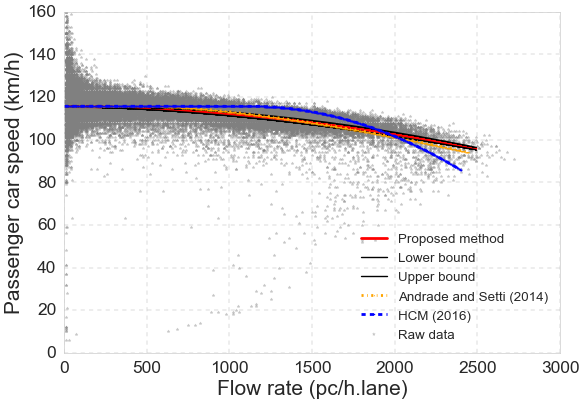}
		\subcaption{}
		\label{fig:sub:subc1}
	\end{minipage}
	\hspace{\fill}
	\begin{minipage}[t]{0.5\textwidth}
		\centering
		\includegraphics[width=0.995\textwidth]{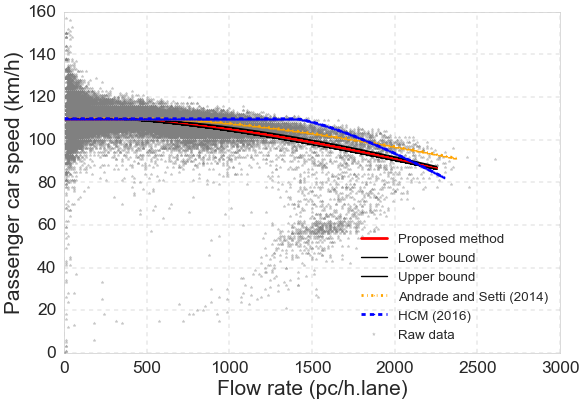}
		\subcaption{}
		\label{fig:sub:subc2}
	\end{minipage}
	\caption{Speed-flow curves for rural highways (\subref{fig:sub:subc1}) Freeway located on SP-348, km 32,0 north and (\subref{fig:sub:subc2}) Multilane located on SP-280 km 51,9 east}
	\label{fig:rural_highways}
\end{figure}

\begin{figure}[!htbp]
	\begin{minipage}[t]{0.5\textwidth}		
		\centering
		\includegraphics[width=0.995\textwidth]{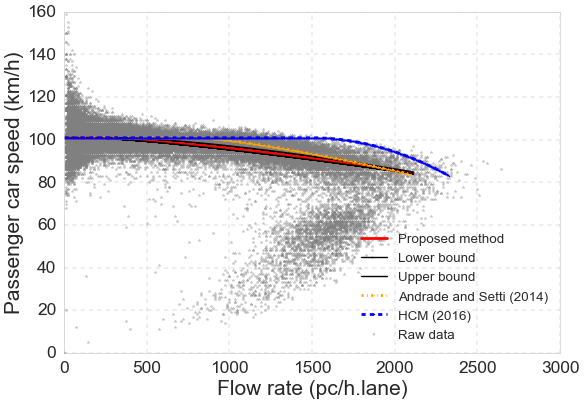}
		\subcaption{}
		\label{fig:sub:subc3}
	\end{minipage}
	\hspace{\fill}
	\begin{minipage}[t]{0.5\textwidth}
		\centering
		\includegraphics[width=0.995\textwidth]{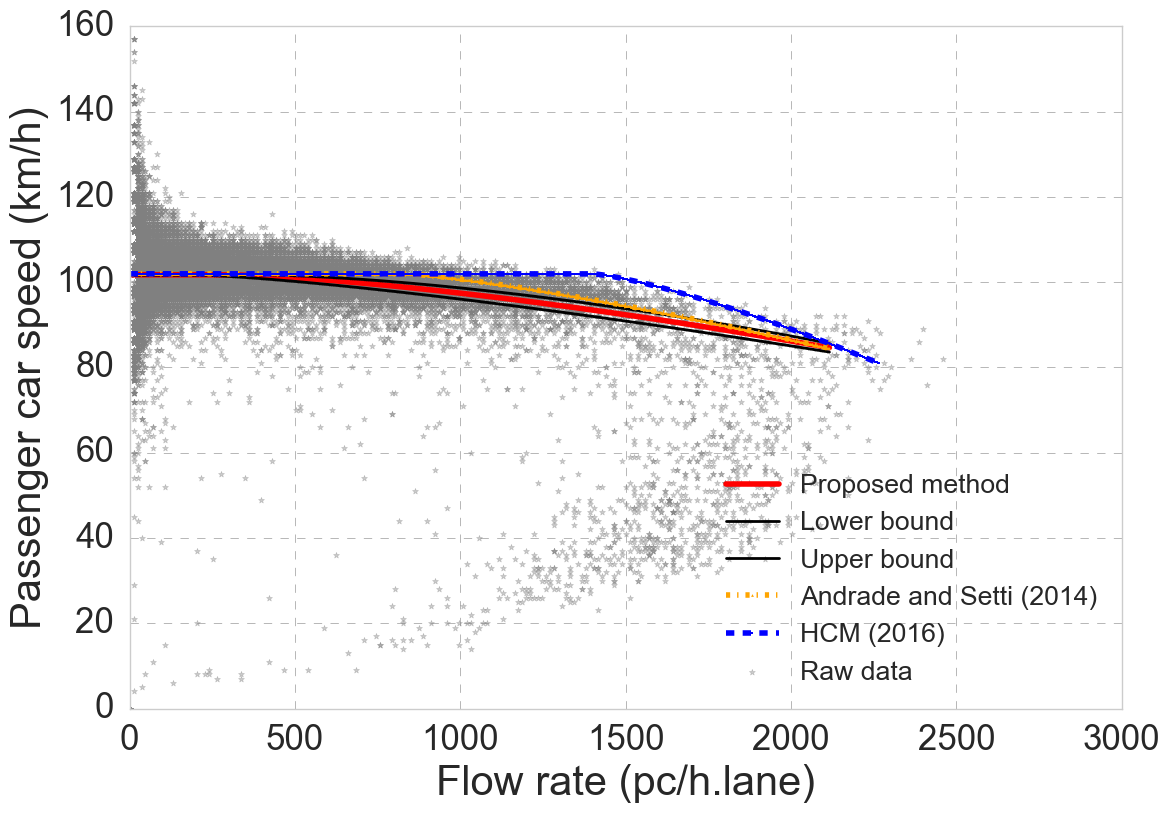}
		\subcaption{}
		\label{fig:sub:subc4}
	\end{minipage}
	\caption{Speed-flow curves for urban highways (\subref{fig:sub:subc3}) Freeway located on SP-280, km 29,5 east and (\subref{fig:sub:subc4}) Multilane located on SP-021, km 18,3 north}
	\label{fig:urban_highways}
\end{figure}

The comparison between methods consists of an analysis of the parameters values estimated by the proposed method, HCM 2016 \cite{hcm2016} and Andrade and Setti \cite{andrade2014}, allied to a graphical analysis of the speed-flow curves. Figures~\ref{fig:rural_highways} and~\ref{fig:urban_highways} show traffic flow and speed observations with the speed-flow curves obtained by the proposed method along with their respective credible regions of 95{\%}, and the curves referring to HCM 2016 \cite{hcm2016} and Andrade and Setti \cite{andrade2014}. 

Figure~\ref{fig:sub:subc1} and~\ref{fig:sub:subc2} represent sensors located, respectively, on SP-348, km 32,0 north and on SP-280, km 51,9 east, which correspond to rural segments of freeway and multilane. Figure~\ref{fig:sub:subc3} and~\ref{fig:sub:subc4} are related to sensors located, respectively, on SP-280, km 29,5 east and on SP-021, km 18,3 north that correspond to urban segments of freeway and multilane. 

In the HCM speed-flow model for multilane and freeways, the curves are defined as a function of the free-flow speed (${\textit{u}}_f$). Thus, the estimation of this parameter becomes a fundamental step for the model calibration. For the HCM 2016 \cite{hcm2016} and Andrade and Setti \cite{andrade2014}, the curves were determined through the free-flow speed, which was calculated by the average speed for successive intervals of traffic flow of 50~pc/(h.lane) to 350~pc/(h.lane) and rounded to the nearest km/h, varying between 72 and 122~km/h.

\begin{figure}[!htbp]
	\centering		
	\begin{minipage}[t]{0.45\textwidth}
		\centering		
		\includegraphics[width=\textwidth]{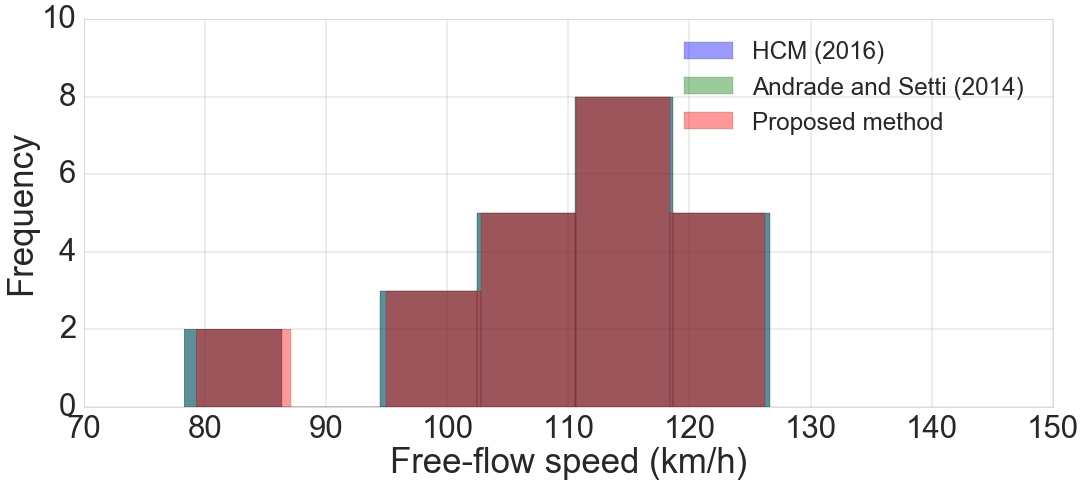}
		\subcaption{}
		\label{fig:sub:subd1}
	\end{minipage}
	\hspace{\fill}
	\begin{minipage}[t]{0.45\textwidth}
		\centering		
		\includegraphics[width=\textwidth]{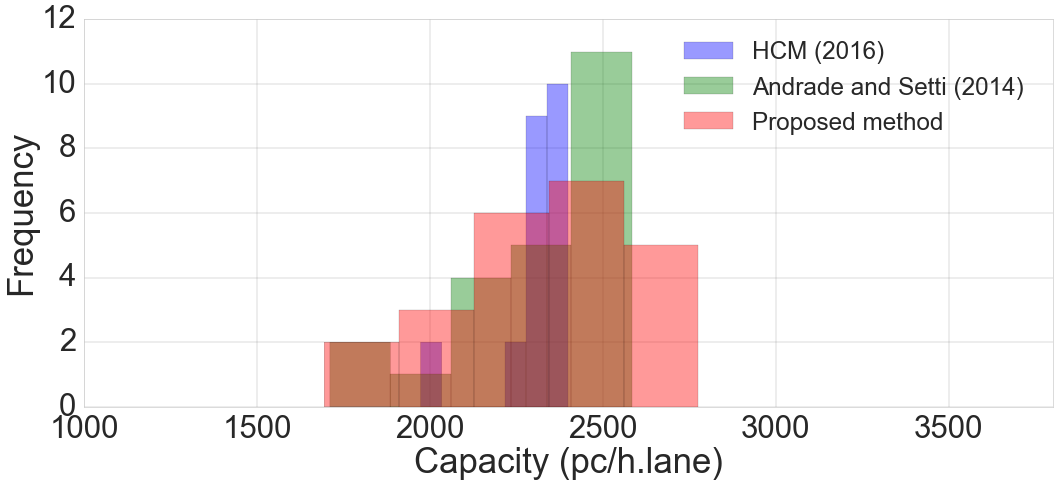}
		\subcaption{}
		\label{fig:sub:subd2}
	\end{minipage}
	\hspace{\fill}
	\begin{minipage}[t]{0.45\textwidth}
		\centering		
		\includegraphics[width=\textwidth]{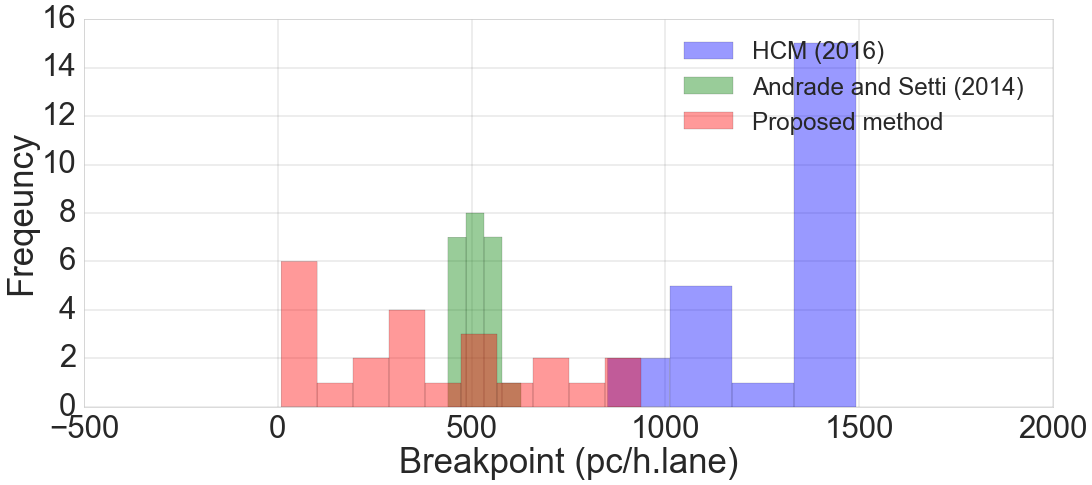}
		\subcaption{}
		\label{fig:sub:subd3}
	\end{minipage}
	\hspace{\fill}
	\begin{minipage}[t]{0.45\textwidth}
		\centering		
		\includegraphics[width=\textwidth]{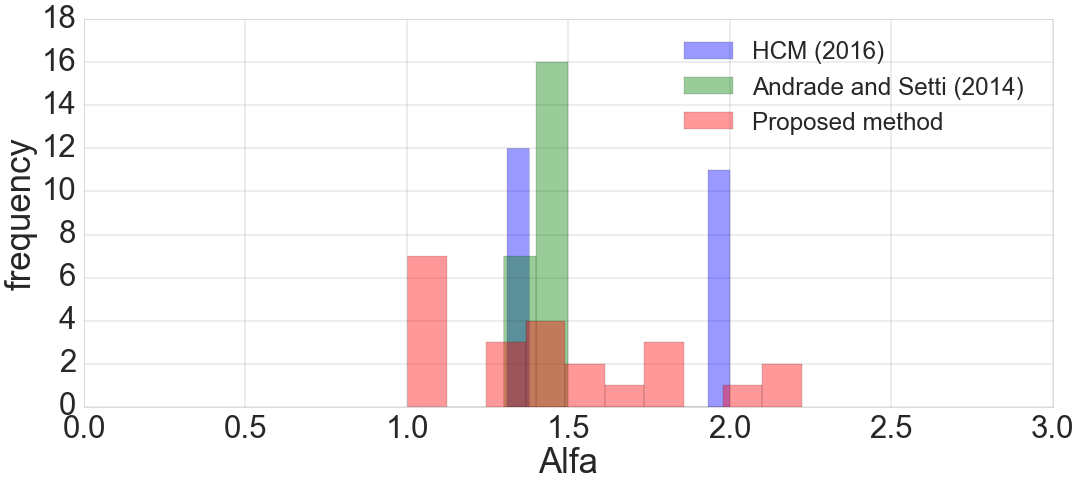}
		\subcaption{}
		\label{fig:sub:subd4}
	\end{minipage}
	\caption{Histograms for the proposed calibration method, HCM 2016 \cite{hcm2016} and Andrade and Setti \cite{andrade2014} for (\subref{fig:sub:subd1}) Free-flow speed (\subref{fig:sub:subd2}) Capacity (\subref{fig:sub:subd3}) Breakpoint and (\subref{fig:sub:subd4}) Calibration coefficient}
	\label{fig:hist_}
\end{figure}

Figure~\ref{fig:hist_} presents the parameters histograms for the proposed calibration method, HCM 2016 \cite{hcm2016} and Andrade and Setti \cite{andrade2014}. In general, the four model parameters, free-flow speed, capacity, breakpoint and calibration coefficient, approximated the values of Andrade and Setti \cite{andrade2014} than the HCM 2016 \cite{hcm2016}, as expected, since it speed-flow curves were adapted to local conditions. 

The free-flow speed obtained by the proposed method presents values really close to the other methods discussed and, therefore, is considered well defined. Regarding to the traffic flow at capacity, the proposed method present values close those obtained by Andrade and Setti \cite{andrade2014} with a maximum difference between then of 205 pc/h.lane and a standard deviation of 108 pc/h.lane. As in comparison to HCM 2016 \cite{hcm2016}, the maximum difference of values is 375 pc/h.lane and a standard deviation of 191 pc/h.lane.

The calibration coefficient obtained by the proposed method comprises a range of values between 1 and 2.22, while the methods used for comparison present only two values, which vary according to the type of highway classification. The average of it values for rural and urban highways is equal to values present in Andrade and Setti \cite{andrade2014}.

\begin{figure}[h]
	\centering
	\includegraphics[width=0.4\textwidth]{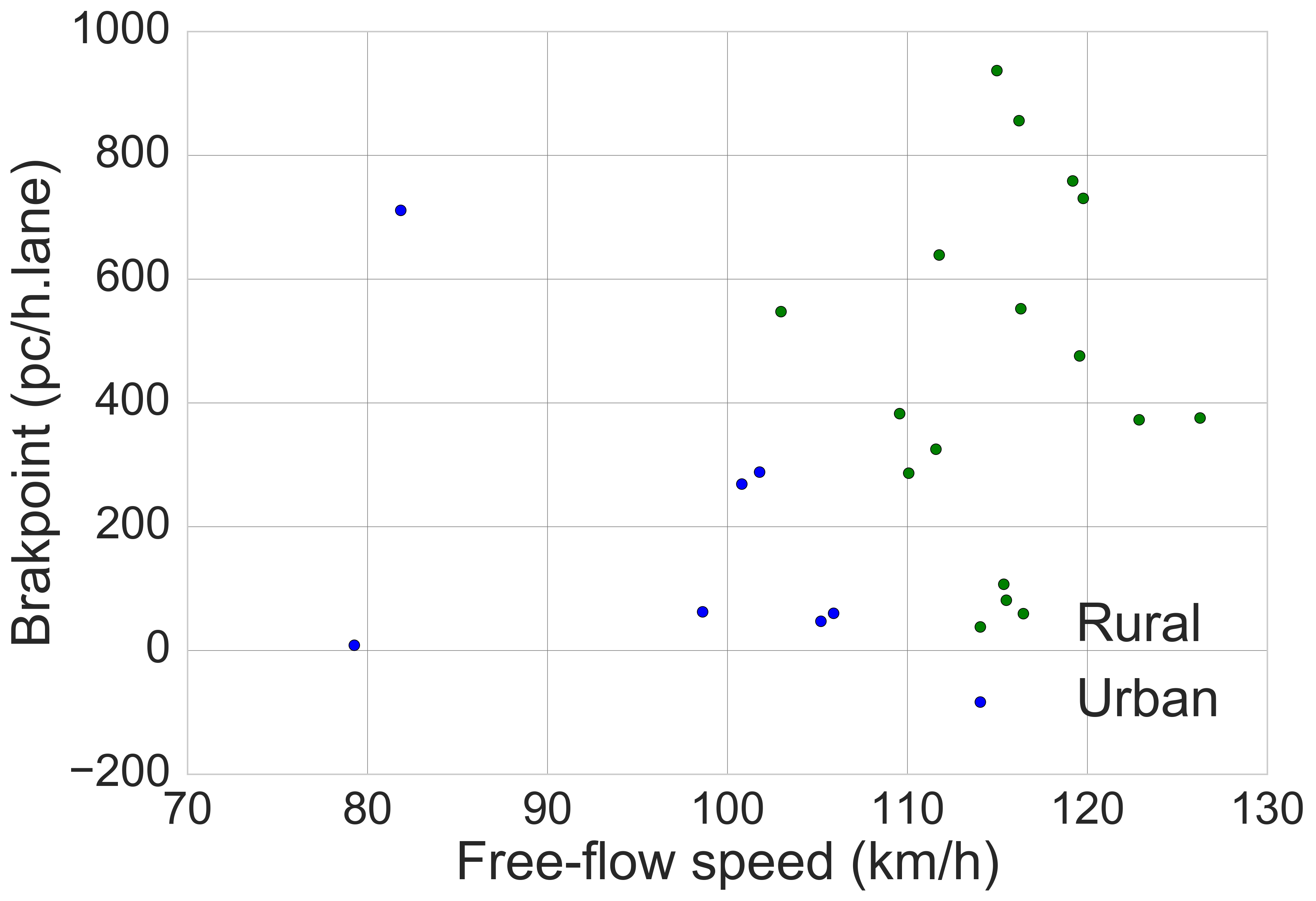}
	\caption{Relation between free-flow speed and breakpoint for rural and urban highways}
	\label{fig:ffs_bp}
\end{figure}

In relation to the breakpoint, this presents values much lower than those obtained by HCM 2016 \cite{hcm2016} and varies around the values of Andrade and Setti \cite{andrade2014}. In some cases the breakpoint obtained by the proposed method is close to zero, which suggests the existence of a plateau in which the free-flow speed remains constant can be much smaller than expected, or even non-existent. Along with that, the breakpoint does not increase with the free-flow speed decrease for freeways or remain constant for multilane as in HCM 2016 \cite{hcm2016}. Figure~\ref{fig:ffs_bp} represents the dispersion graph for urban and rural highways.

As for segments classification, free-flow speed, capacity, speed at capacity, breakpoint, and calibration coefficient tend to be on average higher for rural highways in relation to urban as shown by Table~\ref{tab:ocupacao_lindeira}. This is related to the use of these types of highways, in which rural highways are commonly used for long trips while urbans are related to local trips. 

\begin{table}[htbp]
	\centering
	\caption{Descriptive analysis for abutting land use for the period of one year}
	\resizebox{\columnwidth}{!}{%
		\begin{tabular}{lccccc|ccccc}
			\toprule
			& \multicolumn{10}{c}{Abutting land use} \\
			\cmidrule{2-11}          & \multicolumn{5}{c|}{Rural}            & \multicolumn{5}{c}{Urban} \\
			\cmidrule{2-11}          & ${\textit{u}}_f$   & \multicolumn{1}{l}{${q}_c$} & \multicolumn{1}{l}{\textit{bp}} & \multicolumn{1}{l}{$\alpha$} & \multicolumn{1}{l|}{${\textit{u}}_c$} & ${\textit{u}}_f$   & \multicolumn{1}{l}{${q}_c$} & \multicolumn{1}{l}{\textit{bp}} & \multicolumn{1}{l}{$\alpha$} & \multicolumn{1}{l}{${\textit{u}}_c$} \\
			\cmidrule{2-11}   Count & 16    & 16    & 16    & 16    & 16    & 7    & 7    & 7    & 7    & 7 \\
			Mean & 115,5 & 2.461  & 468 & 1,52  & 91.4  & 96,18  & 2.016  & 206   & 1,31  & 80,6 \\
			Std & 5,6   & 167   & 269   & 0,40  & 10,4   & 11,0  & 209   & 248   & 0,30  & 8,4 \\
			Min   & 103,0 & 2.190  & 59   & 1,01  & 65,1  & 79,3  & 1.694  & 8,7     & 1,00  & 67,7 \\
			25\%  & 111,7 & 2.352  & 316   & 1,24  & 84,5  & 90,2  & 1.900  & 54  & 1,01  & 67,8 \\
			50\%  & 115,8 & 2.489  & 429   & 1,44  & 94,7 & 100,8 & 2.110  & 62.6  & 1,36 & 84,4 \\
			75\%  & 119,3 & 2.583  & 662   & 1,82  & 97,9 & 103,5 & 2.138  & 278  & 1,51  & 85,5 \\
			Max   & 126.3 & 2.776 & 937   & 2,22  & 106,8 & 105,9 & 2.232  & 711   & 1,74  & 89,3 \\
			\bottomrule
		\end{tabular}%
	}
	\label{tab:ocupacao_lindeira}%
\end{table}%

\section{Analysis of temporal variation}

The monthly calibration of the speed-flow relationship of the HCM model was performed for the 23 collection points. For each collection point, it was considered a range of data bigger than 2000 observation per period of analysis - month. This is related to a minimum amount of data considered necessary to perform a calibration and the criteria used to filter the data base - traffic flow only composed by passenger car, traffic under normal operations condition, and traffic under free-flow regime. Thus,  the minimum of month considered appropriated for calibration in a period of one year was 10 and maximum 12. In the results of temporal analysis variation were considered: posterior distributions of parameters and MCMC convergence; parameters variation for each month and comparison with the values obtained with the one year calibration.    

\subsection{Posterior distribution and MCMC convergence}

As in the calibration performed for one year period, all the parameters presented convergence in the MCMC process. For that it was performed a graphical analysis allied to the Gelman and Rubin test. The same MCMC configuration applied for the period of one year was used.

The means were used as representative value for the parameters posterior distribution. The shape of the parameters distribution for all collection point obtained through the monthly calibration presented the same pattern as for the calibration performed in the period of one year. Free-flow speed and capacity presented only Normal distribution while breakpoint and calibration coefficient presented Normal and Exponential distribution.  

\subsection{Parameters estimated monthly}

\begin{figure}[!htbp]
	\begin{minipage}[t]{0.5\textwidth}		
		\centering
		\includegraphics[width=0.995\textwidth]{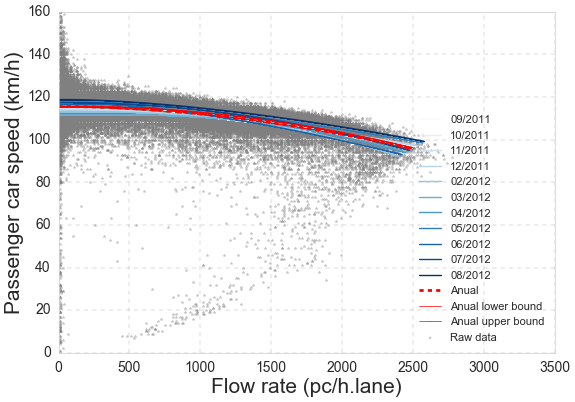}
		\subcaption{}
		\label{fig:sub:sub1}
	\end{minipage}
	\hspace{\fill}
	\begin{minipage}[t]{0.5\textwidth}
		\centering
		\includegraphics[width=0.995\textwidth]{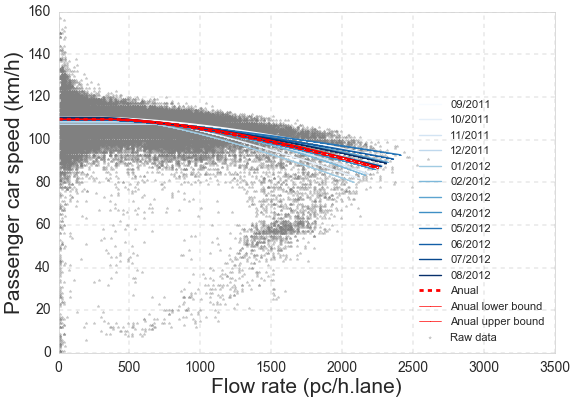}
		\subcaption{}
		\label{fig:sub:sub2}
	\end{minipage}
	\caption{Speed-flow curves for rural highways (\subref{fig:sub:sub1}) Freeway located on SP-348, km 32.0 north and (\subref{fig:sub:sub2}) Multilane located on SP-280 km 51,9 east}
	\label{fig:rural_monthly}
\end{figure}

\begin{figure}[!htbp]
	\begin{minipage}[t]{0.5\textwidth}		
		\centering
		\includegraphics[width=0.995\textwidth]{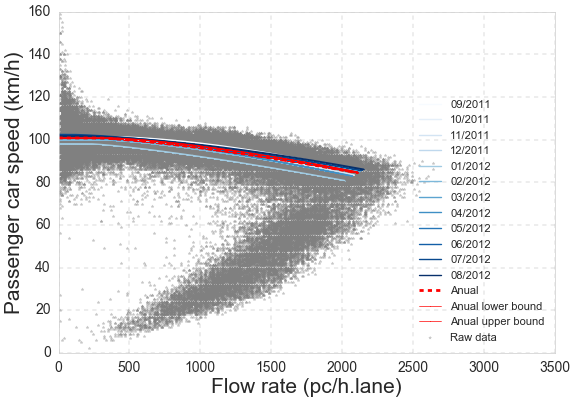}
		\subcaption{}
		\label{fig:sub:sub3}
	\end{minipage}
	\hspace{\fill}
	\begin{minipage}[t]{0.5\textwidth}
		\centering
		\includegraphics[width=0.995\textwidth]{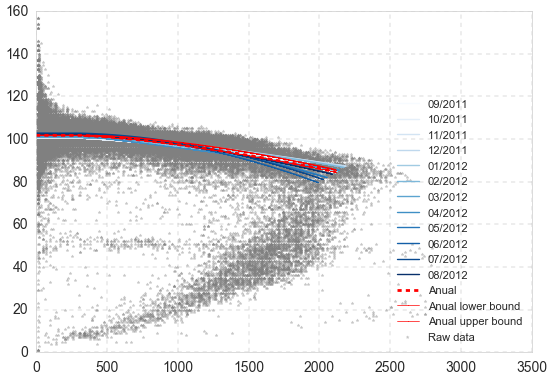}
		\subcaption{}
		\label{fig:sub:sub4}
	\end{minipage}
	\caption{Speed-flow curves for urban highways (\subref{fig:sub:sub3}) Freeway on SP-280, km 29,5 east and (\subref{fig:sub:sub4}) Multilane located on SP-021, km 18,3 north}
	\label{fig:urban_monthly}
\end{figure}

The monthly calibration present parameters with a variance between periods. Figures~\ref{fig:rural_monthly} and~\ref{fig:urban_monthly} show traffic flow and speed observations with the speed-flow curves with their respective credible region of 95{\%} obtained by the calibration for the period of one year along with the curves obtained through the calibration performed monthly. 

\begin{figure*}[!htbp]
	\begin{minipage}[t]{0.5\textwidth}		
		\centering
		\includegraphics[width=0.985\textwidth]{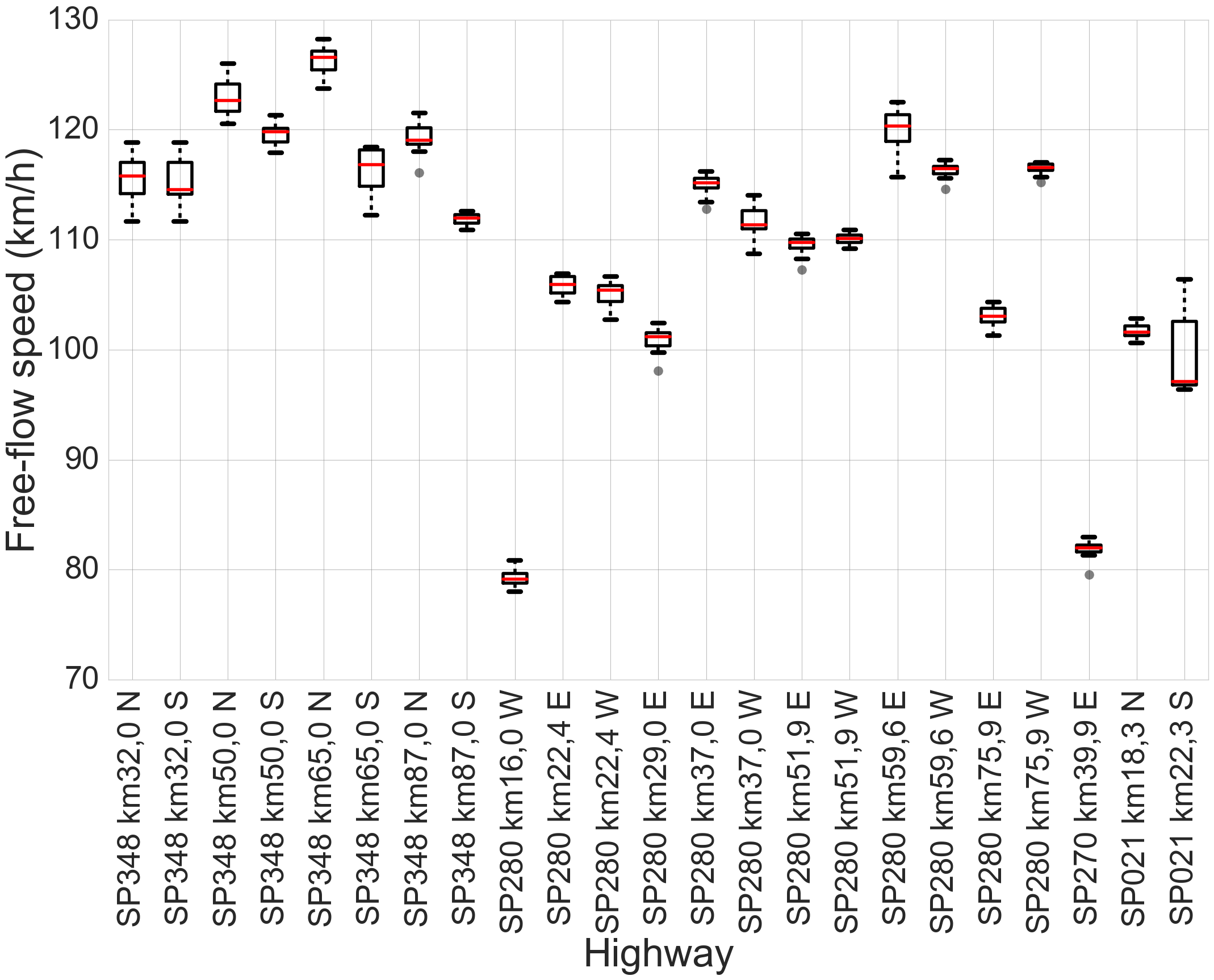}
		\subcaption{}
		\label{fig:sub:subf1}
	\end{minipage}
	\hspace{\fill}
	\begin{minipage}[t]{0.5\textwidth}
		\centering
		\includegraphics[width=0.985\textwidth]{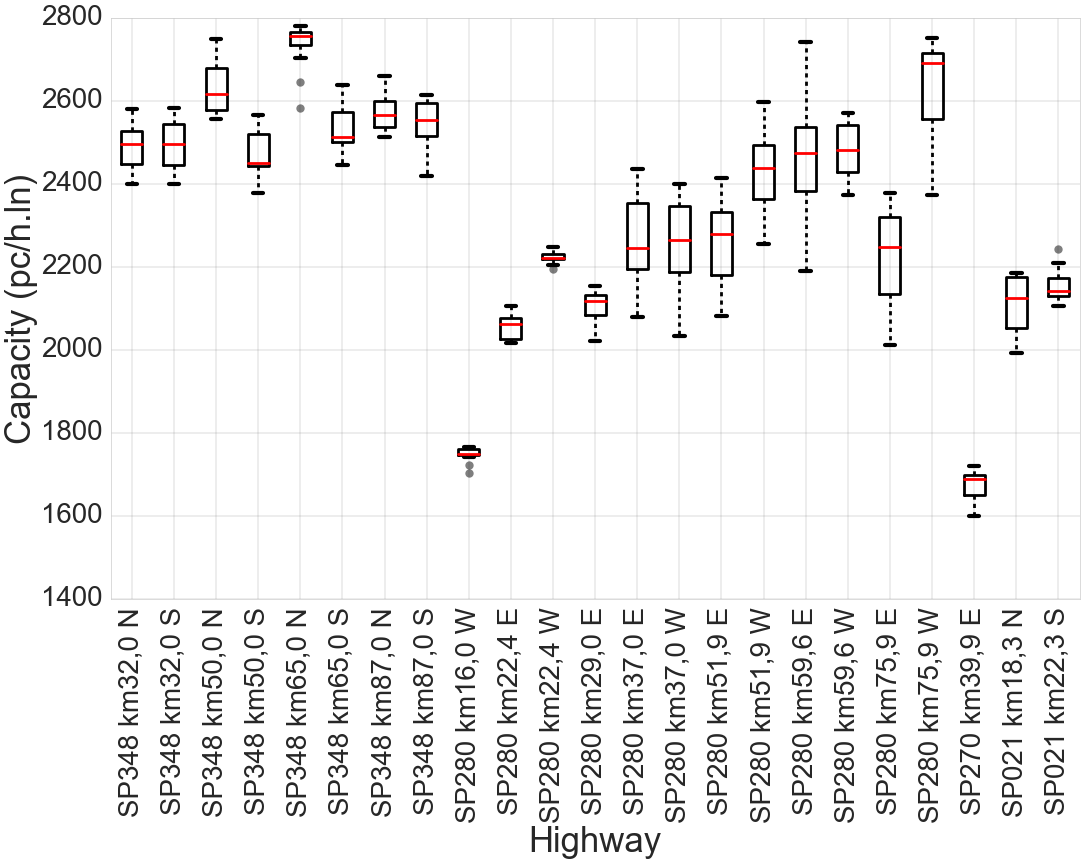}
		\subcaption{}
		\label{fig:sub:subf2}
	\end{minipage}
	\begin{minipage}[t]{0.5\textwidth}
		\centering
		\includegraphics[width=0.985\textwidth]{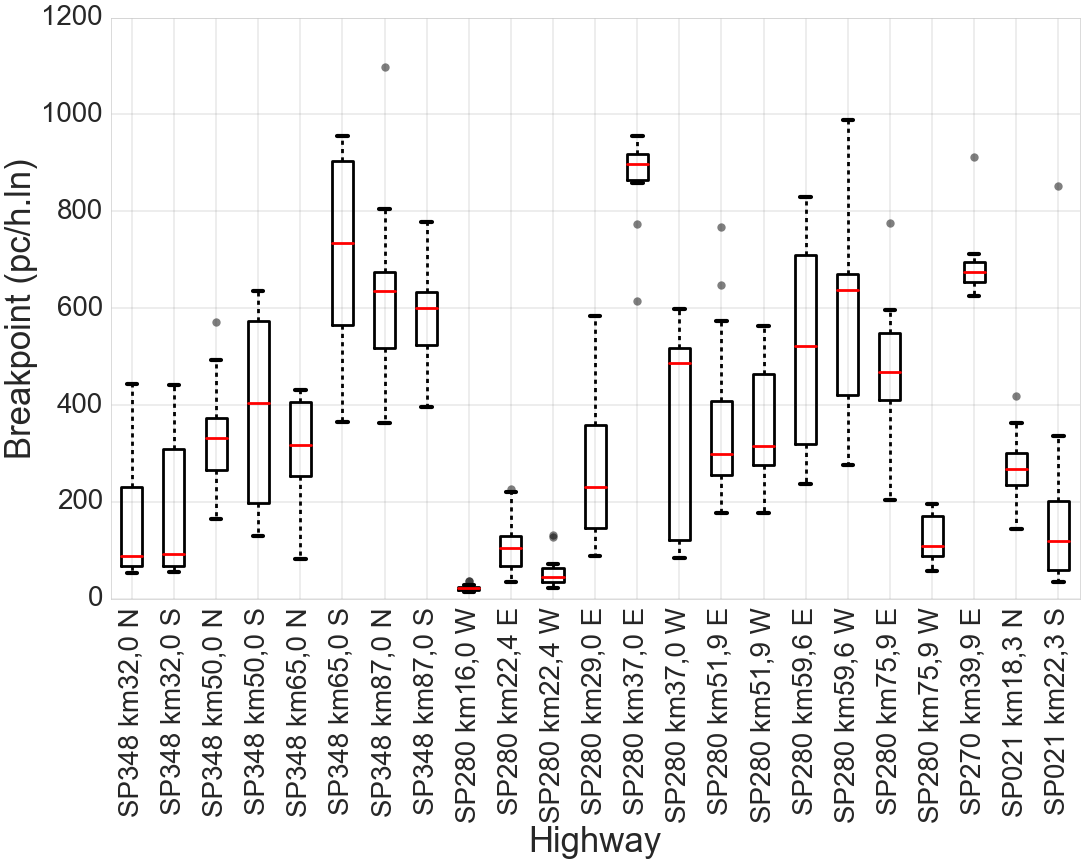}
		\subcaption{}
		\label{fig:sub:subf3}
	\end{minipage}
	\hspace{\fill}
	\begin{minipage}[t]{0.5\textwidth}
		\centering
		\includegraphics[width=0.985\textwidth]{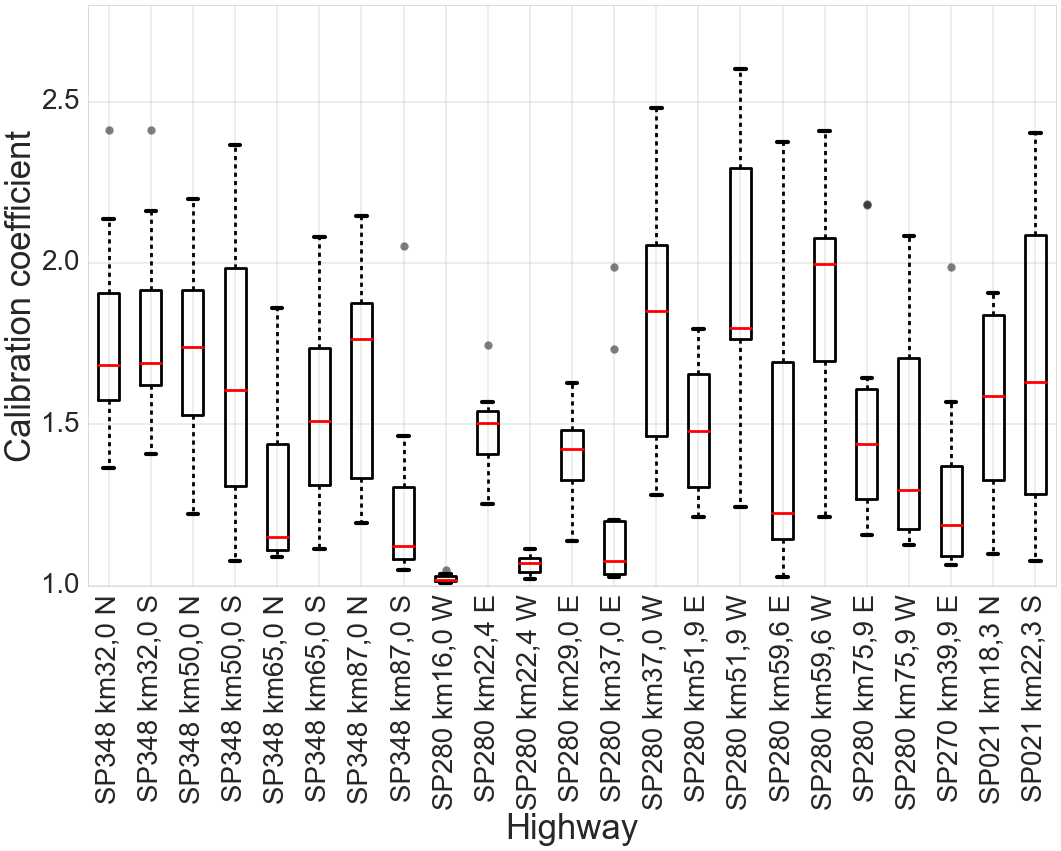}
		\subcaption{}
		\label{fig:sub:subf4}
	\end{minipage}
	\caption{Boxplot graph for (\subref{fig:sub:subf1}) Free-flow speed (\subref{fig:sub:subf2}) Capacity (\subref{fig:sub:subf3}) Breakpoint and (\subref{fig:sub:subf4}) Calibration coefficient}
	\label{fig:boxplot}
\end{figure*}

The curves represent the variation of traffic flow patterns for different periods. This is related to the amount of data used to perform a calibration and to its level of aggregation. Both attributes are directly related to level of detail desired to represent traffic stream. As can been seen, there is a clear variance between months and for the period of one year. Figure~\ref{fig:boxplot} shows the boxplot graph for each parameters estimated monthly for the HCM model calibrated to Brazilian freeways and multilane for all collection point used in the study.

The free-flow speed has different boxplot centers that can be related to the spatial variation of traffic flow which, in this case, is mainly associated with the post speed of each road segment location. The monthly temporal variation is represented for each collection point used in the study where the minimum and maximum standard deviation of 0.54 and 4.13~km/h correspond, respectively, to the sensor located at SP-380 km 87.0 to south and SP021 km 22.3 south. The biggest difference between the minimum and maximum free-flow speed obtained for each month was 10~km/h and is related to the sensor localed on SP-021 km 22,3 south.

For the capacity (${q}_c$) the minimum and maximum standard deviation is equal to 13.7~pc/h.ln and 152.5~pc/h.ln and correspond, respectively to sensor located on SP-280 km 22,4 west and SP-280 km 59,6 east. The biggest difference between the minimum and maximum capacity obtained for each month was 552.2~pc/h.ln and is related to the sensor localed on SP-280 km 59,6 east. 

The breakpoint has the highest number of outliers between all model parameters and presented a minimum standard deviation of 7,3~pc/h.ln and a maximum of 247,5~pc/h.ln related respectively to sensor located on SP-280 km 16,0 west and SP-021 km 22,3 south. Also, the biggest difference of breakpoint for different months into the same collection point was 816,2~pc/h.ln and is related to the sensor localed on SP-021 km 22,3 south. 

Standard deviations from breakpoint and capacity show significant variation. The difference between them in relation to capacity reach up to 63\% for the maximum standard deviation. Noteworthy that both parameters are measured in the same physical quantity, pc/h.lane.

For the calibration coefficient ($\alpha$) the minimum and maximum standard deviation is equal to 0.01 and 0.049 and correspond, respectively to sensor located on SP-280 km 16,0 west and SP-021 km 22,3 south . The biggest difference between the minimum and maximum breakpoint obtained for each month was 1.35 and is related to the sensor localed SP-280 km 51,9 west. 

Overall the free-flow speed presented a small variance of its values for the same collection point and therefore its behavior is considered well defined. The same occurs with the capacity and for that its behavior is also considered well defined. 

Different for the free-flow speed and capacity, the breakpoint presents a high variance for the same collection point and for collections point with similar attributes, as number of lanes, grade and post speed. For that it is considered that breakpoint presented a fuzzy behavior. Lastly, the calibration coefficient has the highest variation in relation to the other traffic parameters and thus is considered to present a behavior not well defined.

\subsection{Relationship between monthly and annual calibration}

\begin{figure}[!htbp]
	\centering		
	\begin{minipage}[t]{0.45\textwidth}
		\centering		
		\includegraphics[width=\textwidth]{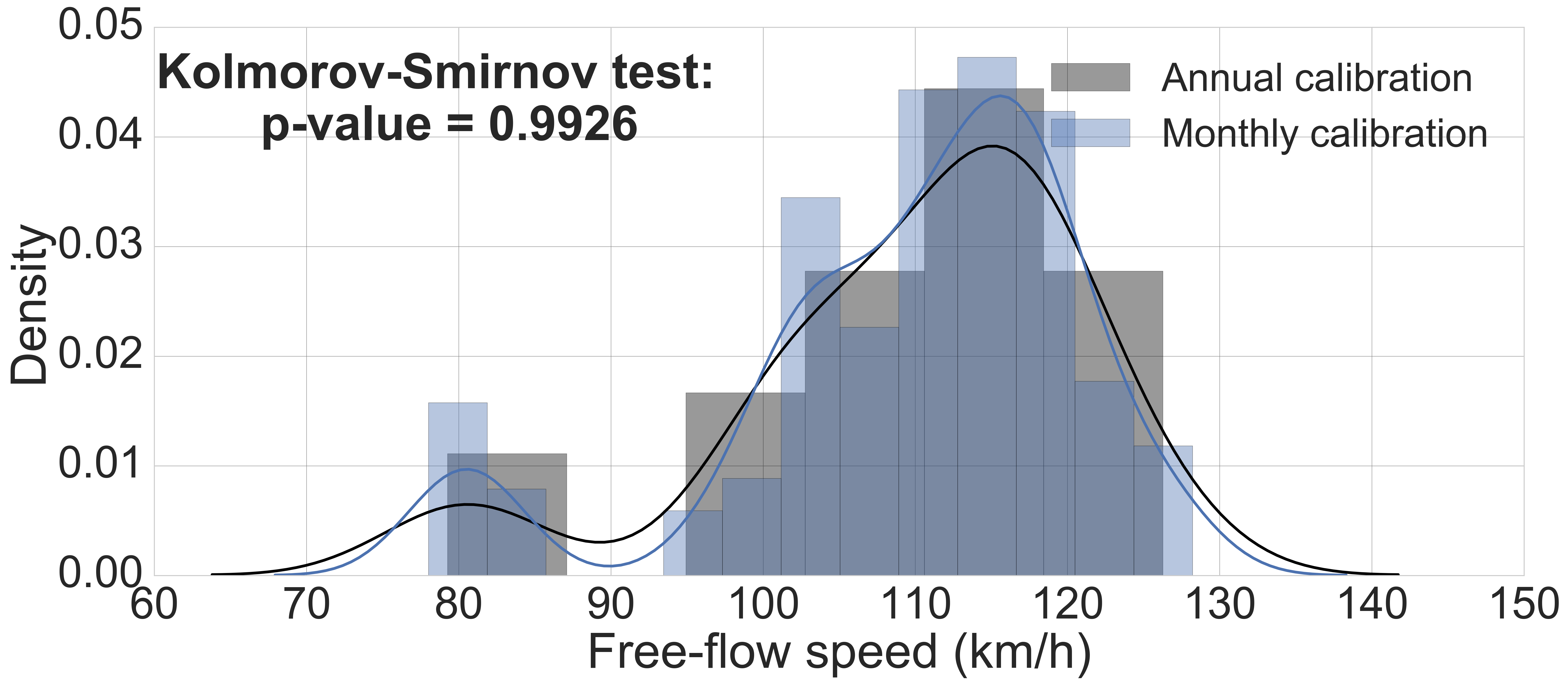}
		\subcaption{}
		\label{fig:sub:subfi1}
	\end{minipage}
	\hspace{\fill}
	\begin{minipage}[t]{0.45\textwidth}
		\centering		
		\includegraphics[width=\textwidth]{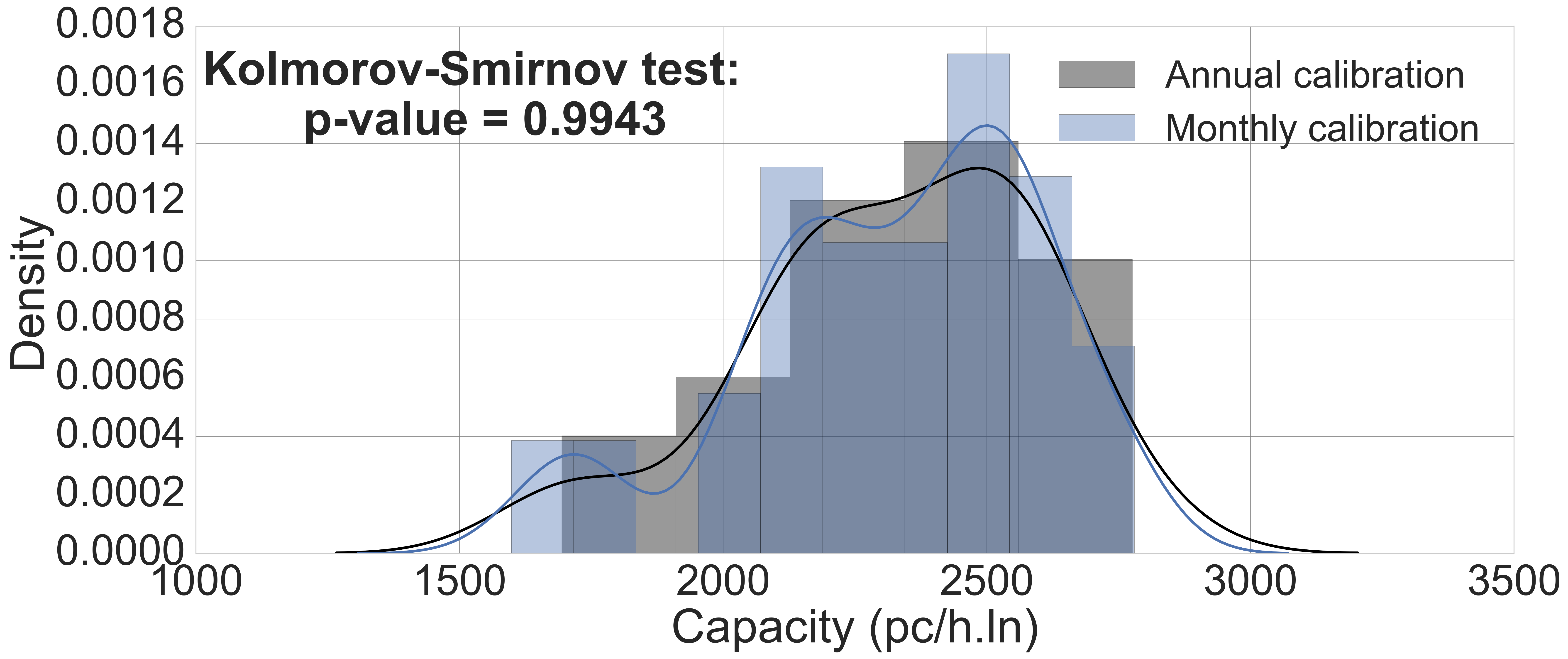}
		\subcaption{}
		\label{fig:sub:subfi2}
	\end{minipage}
	\hspace{\fill}
	\begin{minipage}[t]{0.45\textwidth}
		\centering		
		\includegraphics[width=\textwidth]{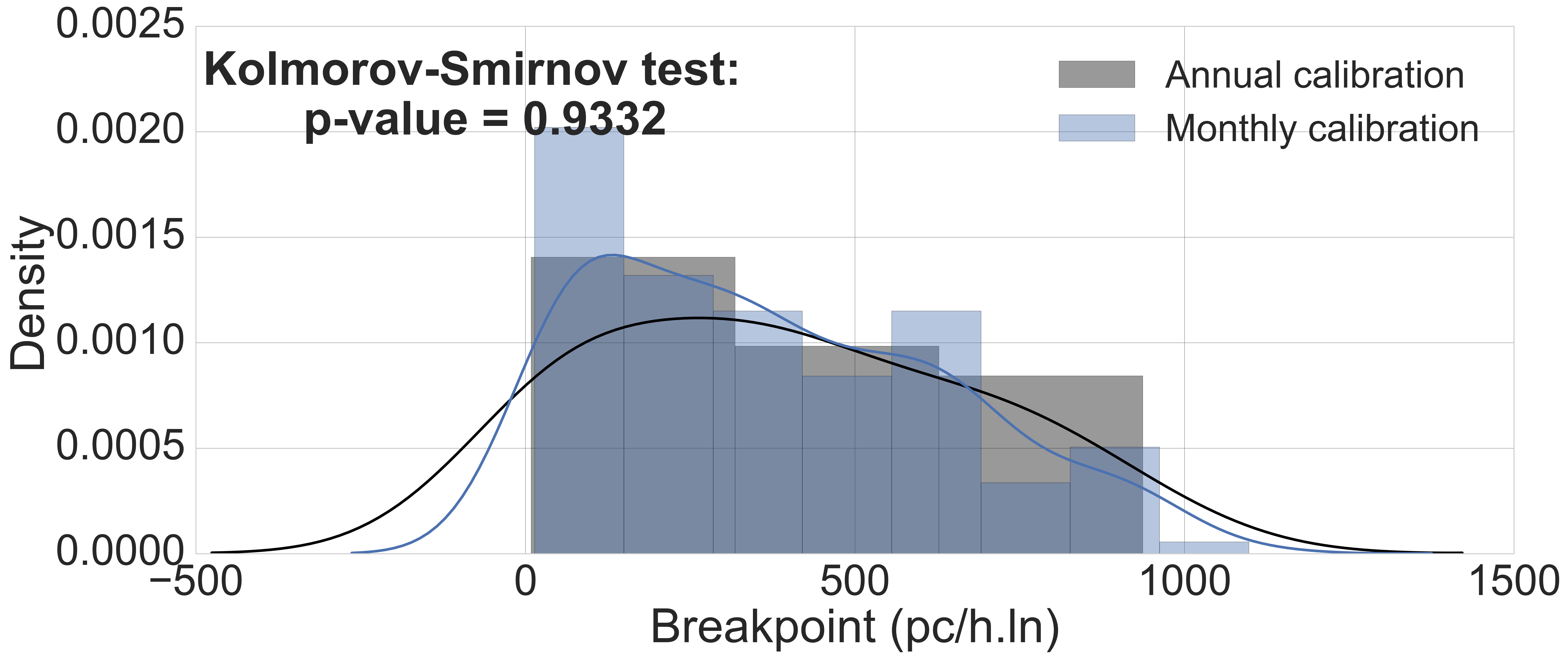}
		\subcaption{}
		\label{fig:sub:subfi3}
	\end{minipage}
	\hspace{\fill}
	\begin{minipage}[t]{0.45\textwidth}
		\centering		
		\includegraphics[width=\textwidth]{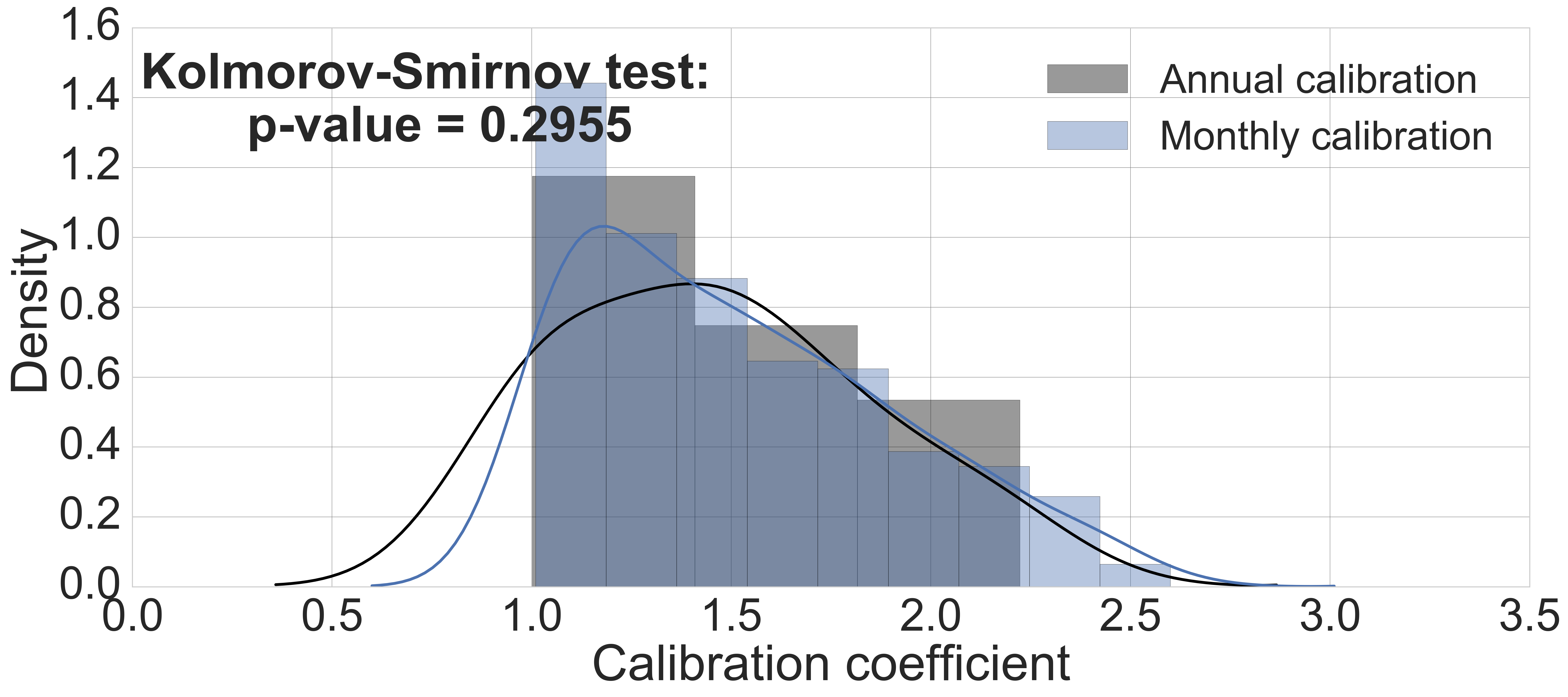}
		\subcaption{}
		\label{fig:sub:subfi4}
	\end{minipage}
	\caption{Probability density function for different periods for (\subref{fig:sub:subfi1}) Free-flow speed (\subref{fig:sub:subfi2}) Capacity (\subref{fig:sub:subfi3}) Breakpoint and (\subref{fig:sub:subfi4}) Calibration coefficient }
	\label{fig:kolgomorov}
\end{figure}

In order to analyze the relationship between the calibration for different periods of time a two-sided Kolmogorov-Smirnov test is performed. The null-hypothesis is that there is no differences between the two distributions. Therefore, it is verified if the traffic parameters distribution obtained monthly and annual are derived from the same probability distribution function. Figure~\ref{fig:kolgomorov} shows for each parameter of traffic the results of the Kolmogorov-Smirnov test and the estimated probability density function obtained for the monthly and annual calibration.

For the free-flow speed, considering a significance level of 0.01, the Kolmogorov-Smirnov test assumes that parameters distribution obtained through monthly and annual calibration are identical as it presents a p-value of 0.9926. The same occurs for capacity as it present a p-value of 0.9943. Different, the breakpoint distributions obtained monthly and annual can no te be considered equal with a significance of 0.01 although it present a high p-value of 0.9322. The calibration coefficient presents a small p-value of 0.2955, which consist that the parameter distribution obtained are not equal. 

\vspace{12pt}
\section{Conclusions and recommendation for future research}

This paper proposes a calibration method for the speed-flow relationship of the HCM 2016 \cite{hcm2016} for freeways and multilane highways. The method is based on Bayesian Inference and was calibrated using speed-flow data collected in 23 permanent traffic sensors in four highways in the state of S{\~a}o Paulo. As advantages, the method provides the models parameters in a unique procedure as a probability distribution function with a credible region, and therefore, speed-flow curve. The metric used to represent the pdf is the distribution mean. As disadvantage, is the method implementation that is more complex than traditional procedures and demands more processing time. 

Compared to the HCM 2016 \cite{hcm2016} and Andrade and Setti \cite{andrade2014}, the proposed calibration method presents: (1) Free-flow speed almost identical; (2) Capacity close to those present on other methods, especially to Andrade and Setti \cite{andrade2014}; (3) Breakpoint ranges values around Andrade and Setti \cite{andrade2014} and significantly lower than on HCM 2016 \cite{hcm2016}, being in some case close to zero, which suggest that traffic stream starts to reduce speed earlier than expected; (4) Also, breakpoint does not increase with the free-flow decrease as described in the HCM 2016 \cite{hcm2016}, instead it presents a fuzzy behavior; (5) calibration coefficient average for rural and urban highways is equal to Andrade and Setti \cite{andrade2014}.

In order to analyze the influence of temporal variation a monthly calibration was performed for all collection points used in the study. The temporal variation represents the monthly traffic stream patterns during a period of one year. The monthly variation presents free-flow speed with a small variance for most of the collection points and in some cases the difference between its values for different months reach up to 10~km/h. Capacity has a relative small variance that goes up to 152~pc/h.lane in the period. The breakpoint shows a high variance for a range of collections points, that could reach up to 250~pc/h.lane while the calibration coefficient presents a high variance for most of the points used in the study.

The free-flow speed and capacity obtained monthly and annual are considered to be derivate from the same probability function as it presented a p-value higher than 0.99 in the two-tail Kolmogorov-Smirnov test. The breakpoint and calibration coefficient obtained monthly and annual cannot be considered from the same probability distribution function considering a significance level of 0.01.

\begin{acknowledgements}
The authors thank the support from ARTESP, CCR RodoAnel, CCR ViaOeste and CCR AutoBAn, which kindly provided traffic data. The authors also acknowledge the financial support provided by CAPES for this research.
\end{acknowledgements}

\bibliographystyle{spphys}       
\bibliography{bibtex}   

\end{document}